\documentclass{99-Styles/MICE}
\usepackage{cite}
\usepackage{textcomp}
\usepackage{times}
\usepackage{bm}         
\usepackage{amsmath,amssymb,bm,slashed} 
\usepackage{amssymb}    
\usepackage{graphicx}   
\usepackage{verbatim}   
\usepackage{color}      
\definecolor{RedViolet}{cmyk}{0.60, 0.99, 0.99, 0.0}
\definecolor{BlueViolet}{cmyk}{0.90, 0.90, 0.0, 0.0}
\definecolor{DarkBlue}{cmyk}{0.90, 0.65, 0.0, 0.0}
\definecolor{DarkGreen}{cmyk}{0.95, 0.25, 0.95, 0.25}
\definecolor{DarkYellow}{cmyk}{0.0, 0.25, 0.95, 0.25}
\usepackage{hyperref}   
\usepackage{longtable}
\usepackage{enumerate}
\usepackage{fmtcount}   

\usepackage{setspace}

\begin{document}

\thispagestyle{empty}

\begin{tabular}{p{0.175\textwidth} p{0.5\textwidth} p{0.225\textwidth}}
    \hspace{-0.8cm}\leftline{~}                                     &
    \centering{~}                                                      &
    \rightline{\today}
\end{tabular}
\vspace{-1.0cm}\\
\noindent{\color{DarkYellow} \rule{\textwidth}{0.43pt}}

\vspace{-0.125cm}
\begin{center}
  {\bf\LARGE\color{RedViolet}
    The Laser-hybrid Accelerator for Radiobiological Applications \\
  }
\end{center}

\makeatletter

\newcommand{\bra}[1]{\ensuremath{\langle #1 |}}   
\newcommand{\ket}[1]{\ensuremath{| #1 \rangle}}   
\newcommand{\bigbra}[1]{\ensuremath{\big\langle #1 \big|}}   
\newcommand{\bigket}[1]{\ensuremath{\big| #1 \big\rangle}}   
\newcommand{\amp}[3]{\ensuremath{\left\langle #1 \,\left|\, #2%
                     \,\right|\, #3 \right\rangle}}  
\newcommand{\sprod}[2]{\ensuremath{\left\langle #1 |%
                     #2 \right\rangle}}  
\newcommand{\ev}[1]{\ensuremath{\left\langle #1 %
                     \right\rangle}} 
\newcommand{\ds}[1]{\ensuremath{\! \frac{d^3#1}{(2\pi)^3 %
                     \sqrt{2 E_\vec{#1}}} \,}} 
\newcommand{\dst}[1]{\ensuremath{\! %
                     \frac{d^4#1}{(2\pi)^4} \,}} 
\newcommand{\tr}{\text{tr}}
\newcommand{\sgn}{\text{sgn}}
\newcommand{\diag}{\text{diag}}
\newcommand{\BR}{\text{BR}}
\newcommand\brabar{\raisebox{-2.0pt}{\scalebox{.2}{\,\,
    \textbf{(}}}\raisebox{-3.25pt}{\scalebox{.8}{{\textendash}}}\raisebox{-2.0pt}
    {\scalebox{.2}{\textbf{)}}}}
\newcommand{\gsim}      {\mbox{\raisebox{-0.4ex}{$\;\stackrel{>}{\scriptstyle \sim}\;$}}}
\newcommand{\lsim}      {\mbox{\raisebox{-0.4ex}{$\;\stackrel{<}{\scriptstyle \sim}\;$}}}

\renewcommand{\vec}[1]{{\mathbf{#1}}}
\renewcommand{\Re}{{\text{Re}}}
\renewcommand{\Im}{{\text{Im}}}
\newcommand{\iso}[2]{{\ensuremath{{}^{#2}}\ensuremath{\rm #1}}}
\newcommand{\eps}{{\ensuremath{\epsilon}}}
\newcommand{\draftnote}[1]{{\bf\color{red} \MakeUppercase{#1}}}
\newcommand{\panm}[1]{{\color{blue} #1}}
\providecommand{\abs}[1]{\lvert#1\rvert}
\providecommand{\norm}[1]{\lVert#1\rVert}

\def\parenbar{\mathpalette\p@renb@r}
\def\p@renb@r#1#2{\vbox{%
  \ifx#1\scriptscriptstyle \dimen@.7em\dimen@ii.2em\else
  \ifx#1\scriptstyle \dimen@.8em\dimen@ii.25em\else
  \dimen@1em\dimen@ii.4em\fi\fi \offinterlineskip
  \ialign{\hfill##\hfill\cr
    \vbox{\hrule width\dimen@ii}\cr
    \noalign{\vskip-.3ex}%
    \hbox to\dimen@{$\mathchar300\hfil\mathchar301$}\cr
    \noalign{\vskip-.3ex}%
    $#1#2$\cr}}}

%
\providecommand{\anmne}{\mbox{$\bar\nu_{\mu} \rightarrow \bar\nu_e$}} 
\providecommand{\nmne}{\mbox{$\nu_{\mu}\rightarrow\nu_e$}} 
\providecommand{\anm}{\mbox{$\bar\nu_\mu$}} 
\providecommand{\nm}{\mbox{$\nu_\mu$}}
\providecommand{\nue}{\mbox{$\nu_e$}} 
\providecommand{\ane}{\mbox{$\bar\nu_e$}} 
\providecommand{\enu}{\mbox{$E_\nu$}}
\providecommand{\piz}{\mbox{$\pi^0 $}}
\providecommand{\pip}{\mbox{$\pi^+$}} 
\providecommand{\pim}{\mbox{$\pi^-$}} 

\setlength{\LTcapwidth}{\linewidth}

\parindent 10pt
\pagestyle{plain}

\vspace{1.5cm}
\begin{center}
  {\bf \color{BlueViolet} The LhARA collaboration} \\
  \vspace{0.25cm}
  G.~Aymar\,$^{1}$,
  T.~Becker\,$^{2}$,
  S.~Boogert\,$^{3}$,
  M.~Borghesi\,$^{4}$,
  R.~Bingham\,$^{5,1}$,
  C.~Brenner\,$^{1}$,
  P.N.~Burrows\,$^{6}$,
  T.~Dascalu\,$^{7}$,
  O.C.~Ettlinger\,$^{8}$,
  S.~Gibson\,$^{3}$,
  T.~Greenshaw\,$^{9}$,
  S.~Gruber\,$^{10}$,
  D.~Gujral\,$^{11}$,
  C.~Hardiman\,$^{11}$,
  J.~Hughes\,$^{9}$,
  W.G.~Jones\,$^{7,20}$,
  K.~Kirkby\,$^{12}$,
  A.~Kurup\,$^{7}$,
  J-B.~Lagrange\,$^{1}$,
  K.~Long\,$^{7,1}$,
  W.~Luk\,$^{7}$,
  J.~Matheson\,$^{1}$,
  P.~McKenna\,$^{5,14}$,
  R.~Mclauchlan\,$^{11}$,
  Z.~Najmudin\,$^{8}$,
  H.T.~Lau\,$^{7}$,
  J.L.~Parsons\,$^{9,21}$,
  J.~Pasternak\,$^{7,1}$,
  J.~Pozimski\,$^{7,1}$,
  K.~Prise\,$^{4}$,
  M.~Puchalska\,$^{13}$,
  P.~Ratoff\,$^{14}$,
  G.~Schettino$^{15,19}$,
  W.~Shields\,$^{3}$,
  S.~Smith\,$^{16}$,
  J.~Thomason\,$^{1}$,
  S.~Towe\,$^{17}$,
  P.~Weightman\,$^{9}$,
  C.~Whyte\,$^{5,14}$,
  R.~Xiao\,$^{18}$
\end{center}
\vspace{2.5cm}
\noindent\textit{\footnotesize
  \begin{tabbing}
    \hspace*{0.45cm}\= \hspace{17.5cm} \kill
     1. \> STFC Rutherford Appleton Laboratory, Harwell Oxford, Didcot, OX11 0QX, UK  \\
     2. \> Maxeler Technologies Limited, 3 Hammersmith Grove, London W6 0ND, UK  \\
     3. \> John Adams Institute, Royal Holloway, University of London, Egham, Surrey, TW20 0EX, UK  \\
     4. \> Queens University Belfast, University Road, Belfast, BT7 1NN, Northern Ireland, UK  \\
     5. \> Department of Physics, SUPA, University of Strathclyde, 16 Richmond Street, Glasgow, G1 1XQ, UK\\
     6. \> John Adams Institute, University of Oxford, Denys Wilkinson Building, Keble Road, Oxford OX1 3RH, UK  \\
     7. \> Imperial College London, Exhibition Road, London, SW7 2AZ, UK  \\
     8. \> John Adams Institute, Imperial College London, Exhibition Road, London, SW7 2AZ, UK  \\
     9. \> University of Liverpool, Liverpool L3 9TA, UK  \\
    10. \> Christian Doppler Laboratory for Medical Radiation Research for Radiation Oncology, Medical University of Vienna, Spitalgasse 23, \\
        \> 1090 Vienna, Austria \\
    11. \> Imperial College NHS Healthcare Trust, The Bays, South Wharf Road, St Mary's Hospital, London W2 1NY, UK  \\
    12. \> University of Manchester, Oxford Road, Manchester, M13 9PL, UK  \\
    13. \> Technische Universit\"at Wien, Atominstitut, Stadionallee 2, 1020 Vienna, Austria \\
    14. \> Cockcroft Institute, Daresbury Laboratory, Sci-Tech Daresbury, Daresbury, Warrington, WA4 4AD, UK  \\
    15. \> National Physical Laboratory, Hampton Road, Teddington, Middlesex, TW11 0LW, UK \\
    16. \> STFC Daresbury Laboratory, Daresbury, Cheshire, WA4 4AD, UK  \\
    17. \> Leo Cancer Care, Broadview, Windmill Hill, Hailsham, East Sussex, BN27 4RY, UK  \\
    18. \> Corerain Technologies, 14F, Changfu Jinmao Building (CFC), Trade-free Zone, Futian District, Shenzhen, Guangdong, China \\
    19. \> University of Surrey, 388 Stag Hill, Guilford, GU2 7XH, UK \\
    20. \> \href{https://www.imperial.ac.uk/cancer-research-uk-imperial-centre/public-and-patients/}{Imperial Patient and Public Involvement Group (IPPIG)}, Imperial College London, Exhibition Road, London, SW7 2AZ, UK \\
    21. \> The Clatterbridge Cancer Centre, Bebington, CH63 4JY, UK \\
    ~   \> \\
    \dag\> Corresponding author, \textit{Email:} a.kurup@imperial.ac.uk\\
  \end{tabbing}
}

\vspace*{\fill}
\leftline{\footnotesize
  Preprint submitted to Frontiers in Physics, Medical Physics and Imaging
}

\newpage
\section*{Abstract}
The `Laser-hybrid Accelerator for Radiobiological Applications',
LhARA, is conceived as a novel, uniquely-flexible facility dedicated
to the study of radiobiology.
The technologies demonstrated in LhARA, which have wide application,
will be developed to allow particle-beam therapy to be delivered in a
completely new regime, combining a variety of ion species in a single
treatment fraction and exploiting ultra-high dose rates.
LhARA will be a hybrid accelerator system in which laser interactions
drive the creation of a large flux of protons or light ions that are
captured using a plasma (Gabor) lens and formed into a beam.
The laser-driven source allows protons and ions to be captured at
energies significantly above those that pertain in conventional
facilities, thus evading the current space-charge limit on the
instantaneous dose rate that can be delivered.
The laser-hybrid approach, therefore, will allow the vast ``terra incognita''
of the radiobiology that determines the response of tissue to ionising
radiation to be studied with protons and light ions using a wide variety
of time structures, spectral distributions, and spatial configurations 
at instantaneous dose rates up to and significantly beyond the ultra-high
dose-rate `FLASH' regime.

It is proposed that LhARA be developed in two stages.
In the first stage, a programme of \emph{in vitro} radiobiology will be 
served with proton beams with energies between 10\,MeV and 15\,MeV.
In stage two, the beam will be accelerated using a fixed-field
accelerator (FFA).
This will allow experiments to be carried out \emph{in vitro} and \emph{in vivo}
with proton beam energies of up to 127\,MeV.
In addition, ion beams with energies up to 33.4\,MeV per nucleon
will be available for \emph{in vitro} and \emph{in vivo} experiments.
This paper presents the conceptual design for LhARA and the R\&D
programme by which the LhARA consortium seeks to establish the
facility.

\section*{Lay summary}

It is well established that radiation therapy (RT) is an effective
treatment for many types of cancer.
Most treatments are delivered by machines that accelerate electrons
which are then used to produce a beam of high-energy photons (X-rays) 
which are directed at a tumour to kill cancer cells.
However, healthy tissue anywhere in the path of the photon beam is
also irradiated and so can be damaged.
Modern X-ray therapy is able to reduce this damage by using several
beams at different angles. 

Recent years have seen the use of a new type of machine in which
protons are accelerated to produce proton beams (rather than photon
beams) which are directed at a tumour.
These proton beams can be arranged to deposit almost all of their
energy in a small volume within a tumour so they cause little damage
to healthy tissue; a major advantage over photon beams. 
But proton machines are large and expensive, so there is a need for
the development of proton machines that are smaller, cheaper and more
flexible in how they can be used. 

The LhARA project is aimed at the development of such proton machines
using a new approach based on high powered lasers.
Such new machines could also make it easier to deliver the dose in
very short high-intensity pulses and as a group of
micro-beams---exciting recent research has shown that this brings
improved effectiveness in killing cancer cells while sparing healthy
tissue.
The technology to be proved in LhARA should enable a course of RT to
be delivered in days rather than weeks and should be more effective. 

Scientifically, there is a need to understand much better the basic
processes by which radiation interacts with biological matter to kill
cancer cells---the investigation of these processes involves physics
as well as biology.
Thus the most important aim of LhARA is to pursue this radiobiological
research in new regimes and from this to develop better treatments.
LhARA will also pursue technological research into laser-hybrid
accelerators.

\newpage
\graphicspath{ {01-Introduction/Figures/} }

\section{Introduction}

Cancer is the second most common cause of death
globally~\cite{WHOwww}.
In 2018, 18.1 million new cancer cases were diagnosed, 9.6
million people died of cancer-related disease, and 43.8 million
people were living with cancer~\cite{Bray2018,Fitzmaurice2018}.
It is estimated that 26.9 million life-years could be saved 
in low- and middle-income countries if radiotherapy capacity could be
scaled up~\cite{ATUN20151153}.
Novel techniques incorporated in facilities that are at once robust,
automated, efficient, and cost-effective are required to deliver the
required scale-up in provision.

Radiation therapy (RT), a cornerstone of cancer treatment, is
used in over 50\% of cancer patients~\cite{DATTA2019918}.
The most frequently used types of radiotherapy employ photon or
electron beams with MeV-scale energies.
Proton and ion beams offer substantial advantages over X-rays
because the bulk of the beam energy is deposited in the Bragg peak.
This allows dose to be conformed to the tumour while sparing healthy
tissue and organs at risk.
The benefits of proton and ion-beam therapy (PBT) are widely
recognised.
PBT today is routinely delivered in fractions of $\sim 2$\,Gy per
day over several weeks; each fraction being delivered at a rate of
$\lsim 10$\,Gy/minute deposited uniformly over the target treatment
volume.
Exciting evidence of therapeutic benefit has recently been reported
when dose is delivered at ultra-high dose-rate, $\gsim 40$\,Gy/s
(``FLASH'' RT) \cite{Favaudon245ra93, Vozenin_2019}, or provided in
multiple micro-beams with diameter less than 1\,mm distributed over
a grid with inter-beam spacing of $\sim 3$\,mm~\cite{Prezado2017}.
However, the radiobiological mechanisms by which the therapeutic 
benefit is generated are not entirely understood.

LhARA, the Laser-hybrid Accelerator for Radiobiological Applications,
is conceived as the new, highly flexible, source of radiation 
that is required to explore the vast ``terra incognita'' of the
mechanisms by which the biological response to ionising radiation is
determined by the physical characteristics of the beam.
A high-power pulsed laser will be used to drive the creation of a
large flux of protons or light ions which are captured and formed into
a beam by strong-focusing plasma lenses. 
The laser-driven source allows protons and ions to be captured at
energies significantly above those that pertain in conventional
facilities, thus evading the current space-charge limit on the
instantaneous dose rate that can be delivered.
The plasma (Gabor) lenses provide the same focusing strength as
high-field solenoids at a fraction of the cost. 
Rapid acceleration will be performed using a fixed-field
alternating-gradient accelerator (FFA) thereby preserving the unique
flexibility in the time, energy, and spatial structure of the beam
afforded by the laser-driven source. 

We propose that LhARA be developed in two stages.
In the first stage, the laser-driven beam, captured and transported
using plasma lenses and bending magnets, will serve a programme of
\emph{in vitro} radiobiology with proton beams of energy of up to 
15\,MeV.
In stage two, the beam will be accelerated using an FFA.
This will allow experiments to be carried out \emph{in vitro} and 
\emph{in vivo} with proton-beam energies of up to 127\,MeV.
Ion beams (including C$^{6+}$) with energies up to 33.4\,MeV per 
nucleon will also be available.

The laser pulse that initiates the production of protons or ions at
LhARA may be triggered at a repetition rate of up to 10\,Hz.
The time structure of the beam may therefore be varied to interrupt
the chemical and biological pathways that determine the biological
response to ionising radiation with 10\,ns to 40\,ns long proton or
ion bunches repeated at intervals as small as 100\,ms. 
The technologies chosen to capture, transport, and accelerate the
beam in LhARA have been made so that this unique capability is
preserved.
The LhARA beam may be used to deliver an almost uniform dose
distribution over a circular area with a maximum diameter of 
between 1\,cm and 3\,cm.
Alternatively the beam can be focused to a spot with diameter of
$\sim 1$\,mm.

The technologies demonstrated in LhARA have the potential to be
developed to make ``best in class'' treatments available to the many.
The laser-hybrid approach will allow radiobiological
studies and eventually radiotherapy to be carried out in completely
new regimes, delivering a variety of ion species in a broad range of
time structures, spectral distributions, and spatial configurations at
instantaneous dose rates up to and potentially significantly beyond
the current ultra-high dose-rate ``FLASH'' regime.

The LhARA consortium is the multidisciplinary collaboration of
clinical oncologists, medical and academic physicists, biologists,
engineers, and industrialists required to deliver such a
transformative particle-beam system.
With its ``pre Conceptual Design Report''
(pre-CDR)~\cite{LhARA:pre-CDR} the consortium lays out its concept
for LhARA, its potential to serve a ground-breaking programme of
radiobiology, and the technological advances that will be made in its
execution.
The work presented in the LhARA pre-CDR lays the foundations for the
development of full conceptual and technical designs for the
facility.
The pre-CDR also contains a description of the R\&D that is required
to demonstrate the feasibility of critical components and systems.
This paper presents a summary of the contents of the pre-CDR and lays
out the vision of the consortium.

\graphicspath{ {02-Motivation/Figures/} }

\section{Motivation}

Conventional (X-ray) RT is one of the most effective cancer treatments, particularly for solid tumours including head and neck tumours and glioblastoma. The dose delivered using X-rays falls approximately exponentially with depth; this characteristic implies a fundamental limit on the maximum dose that can be delivered to the tumour without delivering an unacceptably large dose to healthy tissue. For a given treatment beam-entry point, tumours that lie deep within the patient will receive a dose significantly lower than that delivered to the healthy tissues through which the beam passes on its way to the treatment site. X-rays that pass through the tumour will also deliver a dose to the tissues that lie behind. Dose delivered to healthy tissues can cause the death of the healthy cells and create adverse side effects. Furthermore, the maximum X-ray dose that can be delivered is limited by the presence of sensitive organs such as the brain and spinal cord. This situation is particularly acute in infants for whom dose to healthy tissue, sensitive organs, and bone can lead to developmental issues and a higher probability of secondary malignancies later in life. 

RT delivered using protons and ions, particle-beam therapy (PBT), has the potential to overcome some of the fundamental limitation of X-rays in cancer treatment~\cite{Loeffler2013}. The physics of the interaction between ionising radiation and tissue determines the radiobiological effect. Energy loss through ionisation is the dominant mechanism at the beam energies relevant to PBT. The energy lost per unit distance travelled, the linear energy transfer (LET), increases as the protons or ions slow down. At low velocity, the rate of increase in LET is extremely rapid. This generates a `Bragg peak' in the energy deposited at the maximum range of the beam just as the protons or ions come to rest. In contrast to photons, this characteristic allows the dose delivered to healthy tissue behind the Bragg peak to be reduced to zero for protons, and almost to zero for carbon ions. Scanning the Bragg peak over the tumour volume enables an increase in dose to the tumour while, in comparison to X-ray therapy, sparing tissues in front of the tumour. By choosing carefully the treatment fields, dose to sensitive organs can be reduced significantly compared to an equivalent treatment with photons, thus improving patient outcomes. The Particle Therapy Co-Operative Group (PTCOG) currently lists 90 proton therapy facilities and 12 carbon ion therapy facilities. These facilities are located predominantly in high-income countries~\cite{ptcog}. Low- and middle-income countries (LMIC) are relatively poorly served, indeed nearly 70\% of cancer patients globally do not have access to RT~\cite{DATTA2019918}. Novel RT techniques incorporated in facilities that are robust, automated, efficient, and cost-effective are therefore required to deliver the necessary scale-up in provision. This presents both a challenge and an opportunity; developing the necessary techniques and scaling up RT provision will require significant investment but will also create new markets, drive economic growth through new skills and technologies and deliver impact through improvements in health and well-being. \\

\noindent\textbf{The case for a systematic study of the radiobiology of proton and ion beams} \\
\noindent
The nature of the particle-tissue interaction confers on PBT the advantage that the dose can be precisely controlled and closely conformed to the tumour volume. However, there are significant biological uncertainties in the impact of ionising radiation on living tissue. The efficacy of proton and ion beams is characterised by their relative biological effectiveness (RBE) in comparison to reference photon beams. The treatment-planning software that is in use in the clinic today assumes an RBE value for protons of 1.1~\cite{PAGANETTI201377}. This means that a lower dose of protons is needed to produce the same therapeutic effect that would be obtained using X-rays. However, the rapid rise in the LET at the Bragg peak leads to significant uncertainties in the RBE. It is known that RBE depends strongly on many factors, including particle energy, dose, dose rate, the degree of hypoxia, and tissue type~\cite{Paganetti:2014}, however, the radiobiology that determines these dependencies is not fully understood. A number of studies have shown that there can be significant variation in RBE~\cite{Jones:2018,Giovannini2016,Armin_2018}. Indeed, RBE values from 1.1 to over 3 have been derived from \emph{in vitro} clonogenic-survival assay data following proton irradiation of cultured cell lines derived from different tumours~\cite{Paganetti:2014,Chaudhary2014,Wilkens_2004}. Some of this variation may be due to the positioning of the cells during irradiation relative to the Bragg peak. RBE values of $\sim 3$ are accepted for high-LET carbon-ion irradiation, although higher values have been reported~\cite{Karger_2017}. RBE uncertainties for carbon and other ion species are at least as large as they are for protons.

Uncertainties in RBE can lead to an incorrect estimation of the dose required to treat a particular tumour. Overestimation of the required dose leads to risk of damage to healthy tissue, while an underestimate can lead to the tumour not being treated sufficiently for it to be eradicated. RT causes cell death by causing irreparable damage to the cell’s DNA. Hence, differences in RBE can also affect the spectrum of DNA damage induced within tumour cells. Larger RBE values, corresponding to higher LET, can cause increases in the frequency and complexity of DNA damage, particularly DNA double-strand breaks (DSB) and complex DNA damage (CDD) where multiple DNA lesions are induced in close proximity~\cite{cancers11070946,CARTER2018776}. These DNA lesions are a major contributor to radiation-induced cell death as they represent a significant barrier to the cellular DNA-repair machinery. Furthermore, the specific nature of the DNA damage induced by ions determines the principal DNA-repair pathways employed to effect repair; base excision repair is employed in response to DNA-base damage and single-strand breaks, while non-homologous end-joining and homologous recombination is employed in response to DSBs~\cite{cancers11070946}. However, there are a number of other biological factors that contribute to the efficacy of X-ray therapy and PBT, which produces greatly varying RBE in specific tumours, including the intrinsic radiosensitivity of the tissue, the level of oxygenation (hypoxia), the growth and repopulation characteristics, and the associated tumour micro-environment. Consequently, there is significant uncertainty in the precise radiobiological mechanisms that arise and how these mechanisms are affected by PBT. A more detailed and precise understanding is required for optimal patient-treatment strategies to be devised. Detailed systematic studies of the biophysical effects of the interaction of protons and ions, under different physical conditions, with different tissue types will provide important information on RBE variation and could enable enhanced treatment-planning algorithms to be devised. In addition, studies examining the impact of combination therapies with PBT (e.g. targeting the DNA damage response, hypoxia signalling mechanisms and also the tumour micro-environment) are currently sparse; performing these studies will therefore provide input vital to the development of future personalised patient-therapy strategies using PBT. Such studies are needed, especially in the case of ion-beam radiotherapy. \\

\noindent\textbf{The case for novel beams for radiobiology} \\
\noindent
PBT delivery to date has been restricted to a small number of beam characteristics. In a typical treatment regimen the therapeutic dose is provided in a series of daily sessions delivered over a period of several weeks. Each session consisting of a single fraction of $\sim$2\,Gy delivered at a rate of $\lsim 5$\,Gy/minute. The dose in each fraction would be distributed uniformly over an area of several square centimetres. 
Recent reports provide exciting evidence of therapeutic benefit when the dose is delivered at ultra-high dose rate ($>40$\,Gy/s) ``FLASH'' RT~\cite{Favaudon245ra93, Vozenin_2019}. 
These studies indicate significantly reduced lung fibrosis in mice, skin toxicity in mini-pigs, and reduced side-effects in cats with nasal squamous-cell carcinoma. Varian has indicated that dose rates greater than 40\,Gy/s are useful for FLASH irradiation~\cite{varian_flash}, while IBA have indicated that the FLASH phenomenon is observed at dose rates above 33\,Gy/s~\cite{iba_flash_pr}. In addition, therapeutic benefit has been demonstrated with the use of multiple micro-beams with diameter of less than 1\,mm distributed over a grid with inter-beam spacing of ~3\,mm~\cite{Prezado2017}. However, there is still significant uncertainty regarding the thresholds and the radiobiological mechanisms by which therapeutic benefit is generated in FLASH and micro-beam therapy, which require extensive further study both \emph{in vitro} and in appropriate \emph{in vivo} models.

LhARA is designed to be a highly flexible source delivering the temporal, spectral, and spatial beam structures that are required to elucidate the mechanisms by which the biological response to ionising radiation is determined by the physical characteristics of the beam, including FLASH and micro-beam effects.  These comprehensive studies are not currently possible at clinical RT facilities. Thus the LhARA facility will provide greater accessibility to stable ion beams, enable different temporal fractionation schemes, and deliver reliable and reproducible biological data with fewer constraints than at current clinical centres. The availability of several ion beams (from protons to heavier ions) within the same facility will provide further flexibility and the ability to perform direct radiobiological comparisons of the effect of  different charged particles. In addition, LhARA will enable exhaustive evaluations of RBE using more complex end-points (e.g. angiogenesis and inflammation) in addition to routine survival measurements. The ability to evaluate charged particles in conjunction with other therapies (immunotherapy and chemotherapy), and of performing \emph{in vivo} experiments with the appropriate animal models is a huge advantage given the current lack of evidence in these areas. LhARA therefore has the potential to yield the accumulation of radiobiological data that can drive a signficant change in current clinical practice. The simulations of LhARA that are described in this document have been used to estimate the dose delivered as a function of energy for protons and carbon ions. Details of the simulations can be found in sections~\ref{SubSect:LhARA:invitroBeam} and~\ref{SubSect:LhARA:invivoBeam}. 
The simulations show instantaneous particle rates on the order of 
$10^9$ particles per shot can be achieved, corresponding to average 
dose rates up to $\gsim 120$\,Gy/s for protons and $\gsim 700$\,Gy/s for carbon ions. 
These estimates are based on the baseline specifications for LhARA. \\

\noindent\textbf{Laser-hybrid beams for radiobiology and clinical application} \\
\noindent
High-power lasers have been proposed as an alternative to conventional proton and carbon-ion facilities for radiotherapy~\cite{BULANOV2002240,Fourkal_2003,Malka_2004}. The capability of laser-driven ion beams to generate protons and high-LET carbon ions at FLASH dose rates is a significant step forward for the provision of local tumour control whilst sparing normal tissue. High-power lasers have also been proposed to serve as the basis of electron, proton and ion-beams for radiobiology~\cite{Kraft2010,Fiorini_2011,Doria_2012,Zeil2013,Masood_2014,Zlobinskaya_2014}. More recent projects (e.g. A-SAIL~\cite{A-SAIL}, ELI~\cite{ELI_2013} and SCAPA~\cite{SCAPA_2019}) will also investigate radiobiological effects using laser-driven ion beams. These studies will also address various technological issues~\cite{Manti2017,ROMANO2016153,Masood_2017,Chaudhary_2017,Margarone_2018}.

The LhARA collaboration’s concept is to exploit a laser to drive the creation of a large flux of protons or light ions which are captured and formed into a beam by strong-focusing plasma lenses. The laser-driven source allows protons and ions to be captured at energies significantly above those that pertain in conventional facilities, thus evading the current space-charge limit on the instantaneous dose rate that can be delivered. Rapid acceleration will be performed using a fixed-field alternating-gradient accelerator (FFA) thereby preserving the unique flexibility in the time, energy, and spatial structure of the beam afforded by the laser-driven source.
Modern lasers are capable of delivering a Joule of energy in pulses that are tens of femtoseconds in length at repetition rates of $\gsim10$\,Hz. 
At source, a laser-driven electron beam is reproducibly-well collimated and has a modest ($\sim$5\%) energy spread. 
Laser-driven ion sources create beams that are highly divergent, 
have a large energy spread, and an intensity that can vary by up to
40\% pulse-to-pulse. 
These issues are addressed in the conceptual design through the use
of plasma lenses to provide strong focusing and to allow energy selection.
In addition, sophisticated instrumentation will be used in a fast 
feedback-and-control system to ensure that the dose delivered is both
accurate and reproducible.
This approach will allow produce multiple ion species, from proton to 
carbon, to be produced from a single laser by varying the target foil 
and particle-capture optics. 

The LhARA consortium’s vision is that LhARA will prove the principle of the novel technologies required for the development of future therapy facilities. The legacy of the LhARA programme will therefore be: 
\begin{itemize}
    \item A unique facility dedicated to the development of a deep understanding of the radiobiology of proton and ion beams; and 
    \item The demonstration in operation of technologies that will allow PBT to be delivered in completely new regimes.
\end{itemize}

\graphicspath{ {03-LhARA-facility/Figures/} }

\section{The LhARA facility}
\label{Sect:LhARAFacility}

\graphicspath{ {03-LhARA-facility/03-00-Overview/} }

\begin{figure}
  \begin{center}
    \includegraphics[width=\textwidth]{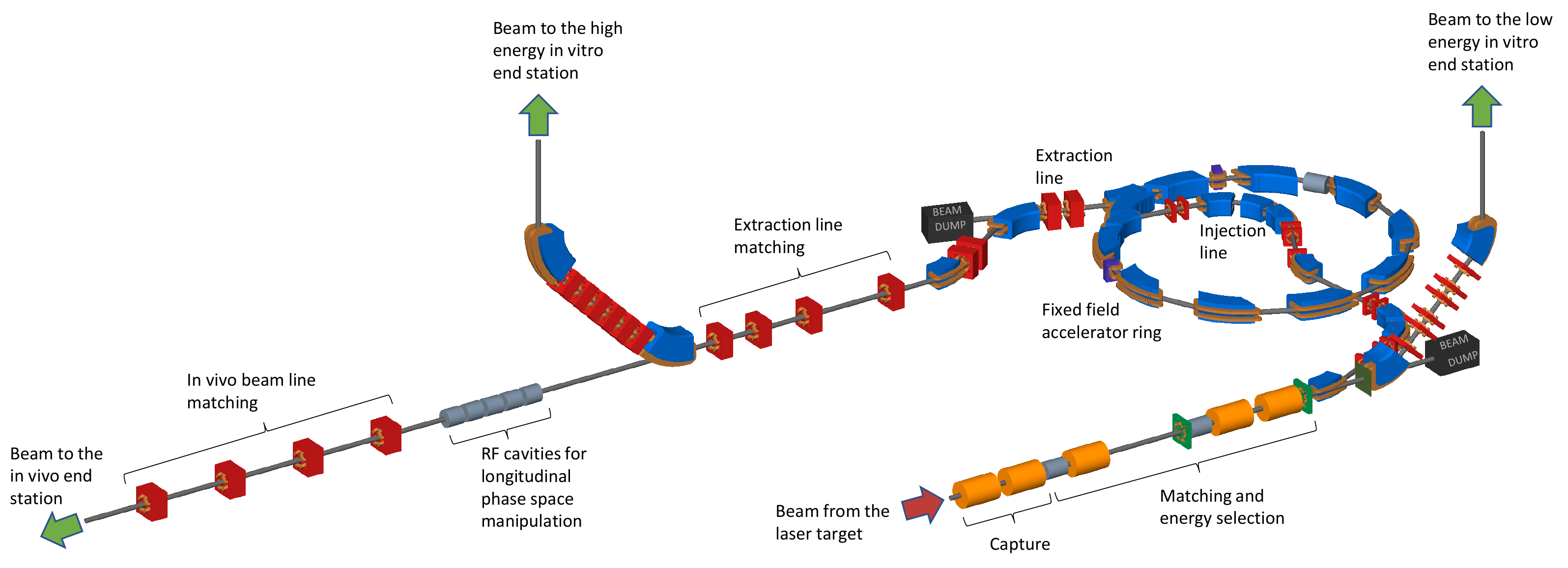}
  \end{center}
  \caption{
    Schematic diagram of the LhARA beam lines.
    The particle flux from the laser-driven source is shown by the red
    arrow.
    The `Capture' section is followed by the `Matching and energy
    selection' section.
    The beam is then directed either into the $90^\circ$ bend that
    takes it to the low-energy \emph{in vitro} end station, towards the FFA
    injection line, or to the low-energy beam dump.
    Post acceleration is performed using the FFA on extraction from which the beam is directed either to the
    high-energy \emph{in vitro} end station, the \emph{in vivo} end station, or the
    high-energy beam dump.
  }
	\label{fig:LhARA-v4.3}
\end{figure}

The LhARA facility, shown schematically in
figure~\ref{fig:LhARA-v4.3}, has been designed to serve two end
stations for \emph{in vitro} radiobiology and one end station for \emph{in vivo} 
studies.
The principle components of the LhARA accelerator are: the laser-driven proton and ion source; the matching and energy selection section; beam delivery to the low-energy \emph{in vitro} end station; the low-energy abort line; the injection line for the fixed-field alternating-gradient accelerator (FFA); the FFA; the extraction line; the high-energy abort line; beam delivery to the high-energy \emph{in vitro} end station; and the transfer line to the \emph{in vivo} end station.
Proton beams with energies of between 12\,MeV and 15\,MeV will be
delivered directly from the laser-driven source to the low-energy
\emph{in vitro} end station via a transfer line.
The high-energy \emph{in vitro} end station and the \emph{in vivo} end station will be
served by proton beams with energy between 15\,MeV and 127\,MeV and by
ion beams, including C$^{6+}$ with energies up to 33.4\,MeV/u.
This configuration makes it natural to propose that LhARA be
constructed in two stages; Stage~1 providing beam to the low-energy
\emph{in vitro} end station and Stage~2 delivering the full functionality of
the facility.
The development of LhARA Stage~1 will include machine performance and
optimisation studies designed to allow \emph{in vitro} experiments to begin
as soon as possible.

The design parameters for the various components of LhARA are given in
tables~\ref{tab:parameters1} and \ref{tab:parameters2}.
The design of the LhARA facility is described in the sections that
follow.
\begin{table}
  \caption{
    Design parameters of the components of the LhARA facility.
    The parameter table is provided in a number of sections.
    This section contains parameters for the Laser-driven proton and
    ion source, the Proton and ion capture section, and the Stage~1
    beam transport section.
  }
  \label{tab:parameters1}
  \begin{center}
    \begin{tabular}{|p{7.5cm}|p{5cm}|p{2cm}|}
      \hline
      {\bf Parameter} & {\bf Value or range} & {\bf Unit}  \\
      \hline
      \multicolumn{3}{|l|}{{\bf Laser driven proton and ion source}} \\
      Laser power   &   100  &   TW  \\
      Laser Energy  &   2.5   &   J   \\
      Laser pulse length   &   25  &   fs  \\
      Laser rep. rate  &   10   &   Hz   \\
      Required maximum proton energy   &   15  &   MeV  \\
      \hline
      \multicolumn{3}{|l|}{{\bf Proton and ion capture}}\\
      Beam divergence to be captured  &   50  &   mrad \\
      Gabor lens effective length & 0.857 & m \\
      Gabor lens length (end-flange to end-flange) & 1.157 & m \\
      Gabor lens cathode radius  & 0.0365 & m \\
      Gabor lens maximum voltage  & 65 & kV \\
      Number of Gabor lenses & 2 & \\
      Alternative technology: solenoid length & 1.157 & m \\
      Alternative technology: solenoid max field strength & 1.3 & T \\
      \hline
      \multicolumn{3}{|l|}{{\bf Stage~1 beam transport}: matching \&
        energy selection, beam delivery to low-energy end station} \\
      Number of Gabor lenses & 3 & \\
      Number of re--bunching cavities & 2 & \\
      Number of collimators for energy selection & 1 & \\
      Arc bending angle & 90 & Degrees \\
      Number of bending magnets & 2 & \\
      Number of quadrupoles in the arc & 6 & \\
      Alternative technology: solenoid length & 1.157 & m \\
a      Alternative technology: solenoid max field strength (to serve the injection line to the Stage~2) & 0.8 (1.4) & T \\
      \hline
    \end{tabular}
  \end{center}
\end{table}
\begin{table}
 \begin{singlespace}
  \caption{
    Design parameters of the components of the LhARA facility.
    The parameter table is provided in a number of sections.
    This section contains parameters for the Stage~2 beam
    transport and the \emph{in vitro} and \emph{in vivo} end stations.
  }
  \label{tab:parameters2}
  \vspace{-0.25cm}
  \begin{center}
    \begin{tabular}{|p{7.5cm}|p{5cm}|p{2cm}|}
      \hline
      {\bf Parameter} & {\bf Value or range} & {\bf Unit}  \\
      \hline
      \multicolumn{3}{|l|}{{\bf Stage~2 beam transport}: FFA, transfer
      line, beam delivery to high-energy end stations} \\
      Number of bending magnets in the injection line & 7 & \\
      Number of quadrupoles in the injection line & 10 & \\
      FFA: Machine type & single spiral scaling FFA & \\
      FFA: Extraction energy & 15--127 & MeV \\
      FFA: Number of cells  & 10  &  \\
      FFA: Orbit R$_{\rm{min}}$  & 2.92 & m \\
      FFA: Orbit R$_{\rm{max}}$  & 3.48 & m \\
      FFA: Orbit excursion & 0.56 & m \\
      FFA: External R & 4 & m \\
      FFA: Number of RF cavities &  2  &  \\
      FFA: RF frequency  & 1.46--6.48 & MHz \\
      FFA: harmonic number & 1, 2 or 4 & \\
      FFA: RF voltage (for 2 cavities) & 4 & kV \\
      FFA: spiral angle & 48.7 & Degrees \\
      FFA: Max B field & 1.4 & T \\
      FFA: k & 5.33 & \\
      FFA: Magnet packing factor & 0.34 & \\
      FFA: Magnet opening angle & 12.24 & degrees \\
      FFA: Magnet gap & 0.047 & m \\
      FFA: Ring tune (x,y) & (2.83,1.22) & \\
      FFA: $\gamma_T$ & 2.516 & \\
      FFA: Number of kickers & 2 & \\
      FFA: Number of septa & 2 & \\
      Number of bending magnets in the extraction line & 2 & \\
      Number of quadrupoles in the extraction line & 8 & \\
      Vertical arc bending angle & 90 & Degrees \\
      Number of bending magnets in the vertical arc & 2 & \\
      Number of quadrupoles in the vertical arc & 6 & \\
      Number of cavities for longitudinal phase space manipulation & 5 & \\
      Number of quadrupoles in the in vivo beam line & 4 & \\
      \hline
      \multicolumn{3}{|l|}{{\bf \emph{In vitro} biological end stations}} \\
      Maximum input beam diameter      &  1-3  & cm \\
      Beam energy spread (full width) & Low-energy end station: $\leq 4$ & \% \\
                                      & High-energy end station: $\leq 1$ & \% \\
      Input beam uniformity & $<5$ & \% \\
      Scintillating fibre layer thickness & 0.25 & mm \\
      Air gap length & 5 & mm \\
      Cell culture plate thickness & 1.3 & mm \\
      Cell layer thickness & 0.03 & mm \\
      Number of end stations &  2  &  \\
      \hline
      \multicolumn{3}{|l|}{{\bf \emph{In vivo} biological end station}} \\
      Maximum input beam diameter      &  1-3  & cm \\
      Beam energy spread (full width)  & $\leq 1$ & \% \\
      Input beam uniformity & $<5$ & \% \\
      Beam options & Spot-scanning, passive scattering, micro-beam &  \\
      \hline
    \end{tabular}
  \end{center}
 \end{singlespace}
\end{table}

\graphicspath{ {03-LhARA-facility/03-01-Ion-source/Figures/} }

\subsection{Laser-driven proton and ion source}
\label{SubSect:LhARA:Src}

\noindent Laser-driven ions have been posited as a source for radiobiological
studies for a number of years \cite{Kraft2010, Yogo2011, Bin2012}.
Until now, the achievable ion energies, energy spreads, and
reproducibility of such beams have meant that such sources are not
suitable for a full radiobiological laboratory setting.
While a number of cell irradiation experiments have been conducted
with laser-accelerated ions \cite{Doria_2012, Zeil2013, Pommarel2017,
Manti2017}, these have been limited in scope to a single-shot
configuration.
In addition, most of these experiments have been performed on
high-power laser facilities with rapidly shifting priorities, where
the time to install dedicated diagnostic systems has not been
available.
At present, a dedicated ion beam for radiobiology, based on a
laser-driven source, is not available anywhere in the world.
Therefore, LhARA will be a unique, state-of-the-art system, able to
explore the radiobiological benefits of a laser-accelerated ion
source.

A novel solution for ion-acceleration is to use a compact,
flexible laser-driven source coupled to a state-of-the-art
beam-transport line.
This allows an accelerating gradient of $\gsim 10$\,GV/m to be
exploited at the laser-driven source.
We propose to operate in a laser-driven sheath-acceleration regime
\cite{Clark2000, Snavely2000, Daido2012} for ion generation.
An intense, short laser pulse will be focused onto a target.
The intense electric field generated on the front surface of the
target accelerates the surface electrons, driving them into the
material.
Electrons which gain sufficient energy traverse the target, ionising
the material as they go.
A strong space-charge electric field, the `sheath', is created as the
accelerated electrons exit the rear surface of the target.
This field in turn accelerates surface-contaminant ions.
The sheath-acceleration scheme has been shown to produce ion energies
greater than 40\,MeV/u at the highest laser intensities.
The maximum proton energy ($E_p$) scales with laser intensity ($I$)
as, $E_p \propto I^\frac{1}{2}$. 
The laser required to deliver a significant proton flux at 15\,MeV can
be compact, relatively inexpensive, and is commercially available. 

The distribution of proton and ion energies observed in laser-driven
beams exhibits a sharp cut off at the maximum energy and,
historically, the flux of laser-accelerated ion beams has varied
significantly shot-to-shot.
To reduce the impact of the shot-to-shot variations the choice has
been made to select particles from the plateau of the two-temperature
energy spectrum of the laser-accelerated ion beam.
This choice should enhance ion-beam stability and allow reproducible
measurements to be carried out at ultra-high dose rates using a small
number of fractions.
To create the flux required in the plateau region it is proposed that
a 100\,TW laser system is used.
A number of commercial lasers are available that are capable of
delivering $>2.5$\,J in pulses of duration $<25$\,fs, at 10\,Hz with
contrast better than $10^{10}:1$. 
Shot-to-shot stability of $<1$\% is promised, an important feature for
stable ion-beam production. 

Key to the operation of this configuration is a system that refreshes
the target material at high-repetition rate in a reproducible manner.
A number of schemes have been proposed for such studies, from
high-pressure gases \cite{Willingale2009,Bin2015,Chen2017}, 
cryogenic hydrogen ribbons
\cite{Margarone-PRX-2016,Gauthier_2017, Obst2017}, liquid sheets 
\cite{Morrison_2018} and tape drives \cite{Noaman-ul-Haq2017}.
For the LhARA facility, a tape drive based on the system developed at
Imperial College London is proposed.
This system is capable of reliable operation at target thicknesses
down to 5\,$\mu$m, using both aluminium and steel foils, and down to
18\,$\mu$m using plastic tapes. 
Such tape-drive targets allow operation at high charge (up to
$100$\,pC at $15 \pm 1$\,MeV, i.e. $>10^9$ protons per shot) and of
delivering high-quality proton and ion fluxes at repetition rates of
up to 10\,Hz or greater.  

The unique features of the laser-driven ion source proposed for LhARA
offer a number of opportunities to push the frontiers in the fields of
sustained high-frequency ion generation, advanced targetry solutions 
and active, high-repetition rate diagnostics.
The successful development and execution of LhARA will provide a leap
forward in terms of capability and open up exciting new opportunities
for applications not just in radiobiology.
While pushing these new frontiers, the radiobiological-capabilities of
LhARA are based on relatively low-energy ion beams, mitigating the
risks that operating at the energy-frontier of the field would imply.

High repetition-rate operation of laser-driven radiation sources is a
relatively new area of interest \cite{Noaman-ul-Haq2017,
Aurand2019, Streeter2018, Dann2019, Kirschner2019}.
Such operating schemes pose a number of engineering challenges.
It is proposed to apply machine-learning and genetic algorithms to the
optimisation of the laser-target interaction to optimise the beam
charge, peak energy, energy spread, and divergence of the ion flux
produced \cite{Aurand2019}.
These techniques will require appropriate R\&D effort.
The first experiments of this kind will be possible using the
existing laser capabilities at Imperial College London, the Central
Laser Facility at the  Rutherford Appleton Laboratory, and elsewhere.

The careful control of the tension on the tape in a tape-drive target
is critical for reproducible operation.
The tape must be stretched to flatten the surface, without stretching
it to its plastic response.
Surface flatness is important for a number of reasons.
Rippling of the front surface modifies the laser absorption
dramatically; uncharacterised rippling can make shot-to-shot 
variations significant and unpredictable \cite{Noaman-ul-Haq2017}.
Similarly, rear surface perturbations can modify the sheath field,
resulting in spatial non-uniformities of the proton beam or
suppression of the achievable peak energies.
Tape drives with torsion control and monitoring to maintain a
high-quality tape surface have been designed and operated in
experiments at Imperial College London.
The development of these targets continues with a view to the
production of new, thinner tapes for improved ion generation
and the creation of ion species other than proton and carbon.
This is an active area of R\&D that will continue with the development
of LhARA.

High repetition-rate ion-beam diagnostics will also need to be
developed.
Such diagnostics will need to measure both the energy spectrum and the
spatial profile of the beams.
Current methods are destructive and are often limited to
low-repetition rate.
Passive detectors have not been demonstrated in the conditions that
will pertain at LhARA.
Technologies being evaluated to address the issues raised by
ion-source diagnostics for LhARA are discussed in
section~\ref{SubSect:LhARAFac:Instr}. 

\graphicspath{ {03-LhARA-facility/03-02-Capture/} }

\subsection{Proton and ion capture}
\label{SubSect:LhARA:PnIonCapture}

The use of an electron cloud as a focusing element for charged-particle beams was first proposed by Gabor in 1947 \cite{GABOR1947}. Gabor noted that a cloud of electrons uniformly distributed about the axis of a cylindrical vessel would produce an ideal focusing force on a beam of positively charged particles. The focal length of such a lens scales with the energy of the incoming particle beam allowing such lenses to provide strong focussing of high-energy beams. Confinement conditions in the radial and axial directions can be determined \cite{pozimski_aslaninejad_2013}.
In the radial direction, where there is magnetic confinement and Brillouin flow, the number density of electrons, $n_e$, that can be contained is given by:
\begin{equation}
    n_{e} = \frac{\epsilon_{0}B^2}{2 m_e} \,; 
\end{equation}
where $B$ is the magnetic field, $m_e$ the mass of the electron, and $\epsilon_0$ the permittivity of free space.
In the longitudinal direction there is electrostatic confinement for which $n_e$ is given by:
\begin{equation}
    n_{e} = \frac{4\epsilon_{0}V_{A}}{eR^{2}} \,; 
\end{equation}
where $e$ the magnitude of the charge on the electron and $R$ is the radius of the cylindrical anode which is held at the positive potential $V_A$. For the electron densities of interest for LhARA the required anode voltage is of the order of 50\,kV. 

In the thin lens approximation, the focal length, $f$, of a Gabor lens can be expressed in terms of the magnetic field and the particle velocity, $v_p$ \cite{72912}:
\begin{equation}
    \frac{1}{f}=\frac{e^{2}B^{2}}{4m_{e}m_{p}v_p^2}l \,;
\end{equation}
where $m_{p}$ is the mass of the particles in the beam. 
The focal length of the Gabor lens is therefore proportional to the kinetic energy or, equivalently, the square of the momentum, of the incoming beam. By comparison, the focal length for a solenoid is proportional to the square of the momentum and that of a quadrupole is proportional to momentum. At the particle energies relevant to LhARA the Gabor lens, or the solenoid, is therefore preferred.

An expression for the focal length as a function of electron number density can be derived by substituting equation (1) into equation (3) to give:
\begin{equation}
    \frac{1}{f}=\frac{e^{2}n_{e}}{4\epsilon_{0}U}l \,;
\end{equation}
where $U = \frac{1}{2}m_{p}v_{p}^2$ is the kinetic energy of the particle beam. 
The focal length of the Gabor lens is inversely proportional to the number density of electrons trapped in the cloud. The focal lengths desired to capture the proton and ion beams at LhARA have been chosen such that the required electron number densities are conservative and lie within the range covered in published experiments. 

For a given focal length, the magnetic field required in the Gabor lens 
is reduced compared to that of a solenoid that would give equivalent 
focusing.
In the non-relativistic approximation the relationship between the magnetic
field in the Gabor lens, $B_{\rm GBL}$, and the equivalent solenoid, 
$B_{\rm sol}$, is given by \cite{pozimski_aslaninejad_2013}:
\begin{equation}
    B_{GPL}=B_{sol}\sqrt{Z\frac{m_e}{m_{ion}}} \, ;
\end{equation}
where $m_{ion}$ is the mass of the ions being focused, and $Z$ is the 
charge state of the ions.
In the case of a proton beam the reduction factor is 43. 
Thus, for example, where a 2\,T superconducting solenoid would be required, 
the magnetic field required for a Gabor lens would only be 47\,mT.
This means the cost of the solenoid for a Gabor lens can be significantly 
lower than the cost for a solenoid of equivalent focusing strength.

\begin{figure}
  \begin{center}
    \includegraphics[width=0.6\textwidth]{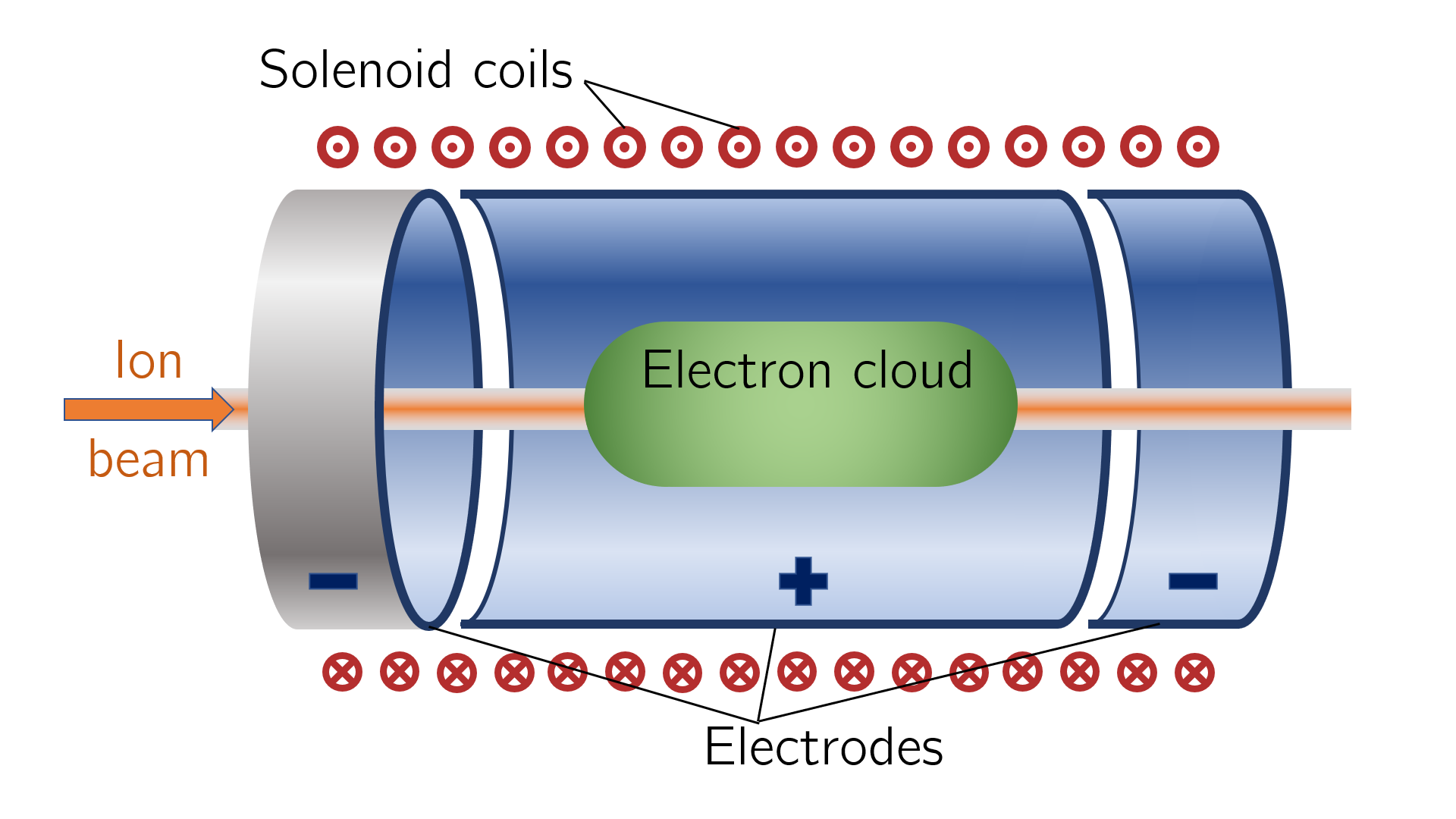}
  \end{center}
  \caption{
    Schematic diagram of a Penning-Malmberg trap of the type proposed
    for use in the Gabor lenses to be used in LhARA.
    The solenoid coils, and the direction of current flow, are
    indicated by the red circles.
    The confining electrostatic potential is provided using a
    central cylindrical anode and two cylindrical negative end
    electrodes.
    The ion beam enters on-axis from the left and the electron cloud
    is indicated by the green shaded area.
  }
  \label{fig:Penning_trap}
\end{figure}

The plasma-confinement system described above is commonly known as a
`Penning trap' and has found wide application in many
fields~\cite{Thompson_penn_trap}. 
Variations on the Penning trap where axial apertures in the cathodes
are introduced, such as the Penning-Malmberg
trap~\cite{Malmberg1980,Malmberg1988} are attractive for beam-based
applications due to the excellent access provided to the plasma
column, see figure~\ref{fig:Penning_trap}. 

Instability of the electron cloud is a concern in the experimental 
operation of Gabor lens; azimuthal beam disruption due to the 
diocotron instability has been observed and described 
theoretically~\cite{Meusel:2013eya}. 
Theory indicates that the diocotron instability is most problematic under well-defined geometric conditions. The reliable operation of a Gabor lens in a regime free from this instability has yet to be demonstrated.
Gabor lenses promise very strong focusing, simple construction, and low magnetic field, all attractive features for LhARA. However, these attractive features come at the cost of relatively high voltage operation ($\gsim 50$\,kV) and possible vulnerability to instability. 

With reliable operation of Gabor lenses as yet unproven, we plan a two-part experimental and theoretical programme of research to prove Gabor-lens suitability. Initial work will include: theoretical investigation of lens stability in a full 3D particle-in-cell code such as VSIM \cite{VSIM}; and the development of electron-density diagnostics based on interferometric measurement of the refractive-index change. These activities will be applied to a time-invariant electron cloud. 
A test Gabor lens will be constructed to allow validation of both the simulation results and a new diagnostic using an alpha emitter as a proxy for the LhARA beam. In addition, the initial investigation will include the design of an electron beam to fill the lens. This last objective will enable the second part of the experimental project; the operation of the Gabor lens in short pulses. It is attractive to match the timing of the establishment of the electron cloud within the Gabor lens to that of the beam and thereby limit instability growth.  
The research project is time limited such that, should it not prove possible to produce a suitable Gabor lens, there will remain time sufficient to procure conventional solenoids in their place.

\graphicspath{ {03-LhARA-facility/03-03-In-vitro-beam-transport/Figures/} }

\subsection{Beam transport and delivery to the low-energy \emph{in vitro} end station}
\label{SubSect:LhARA:invitroBeam}

\noindent Beam-transport from the laser-driven ion source and delivery to the
low-energy \emph{in vitro} end station is required to deliver a uniform dose
distribution at the cell layer.
Beam losses must be minimised for radiation safety and to maximise the
dose that can be delivered in a single shot.
The transport line has been designed to minimise regions in which the
beam is brought to a focus to reduce the impact of space-charge forces
on the beam phase-space.
An optical solution was initially developed using
Beamoptics \cite{beamoptics} and MADX~\cite{Grote:2003ct}.
Accurate estimation of the performance of the beam line requires the
inclusion of space-charge forces and particle-matter interactions.
Therefore, performance estimation was performed using Monte Carlo
particle-tracking from the ion source to the end station.
BDSIM \cite{bdsim}, which is based on the Geant4 toolkit was used
for the simulation of energy deposition arising from beam interactions
with the material in the accelerator and the end station. 
GPT \cite{DeLoos:860825} was used for evaluating the full 3D impact of space-charge.

An idealised Gaussian beam was generated with a spot size of 4\,$\mu$m
FWHM, an angular divergence of 50\,mrad, 35\,fs FWHM bunch length, and
an energy spread of $1 \times 10^{-6}$\,MeV.
The maximum estimated bunch charge is $1 \times 10^{9}$ protons.
The presence of a substantial electron flux produced from the laser
target compensates the high proton charge density in the vicinity of
the ion-production point.
Therefore, the first 5\,cm of beam propagation was simulated without
space-charge.
Beyond this, the proton beam will have separated from the lower energy
electrons sufficiently for space-charge to become a prominent effect 
and cause an emittance growth.
Therefore, a further 5\,cm drift was simulated including space-charge
forces.
At a distance of 10\,cm from the ion source the beam is at the exit
of the laser-target vessel.
The kinematic distributions of ions in the beam were stored at this
point and passed to the relevant BDSIM and GPT simulations of the
downstream beam line.  

\begin{figure}
	\begin{center}
		\includegraphics[width=0.9\textwidth]{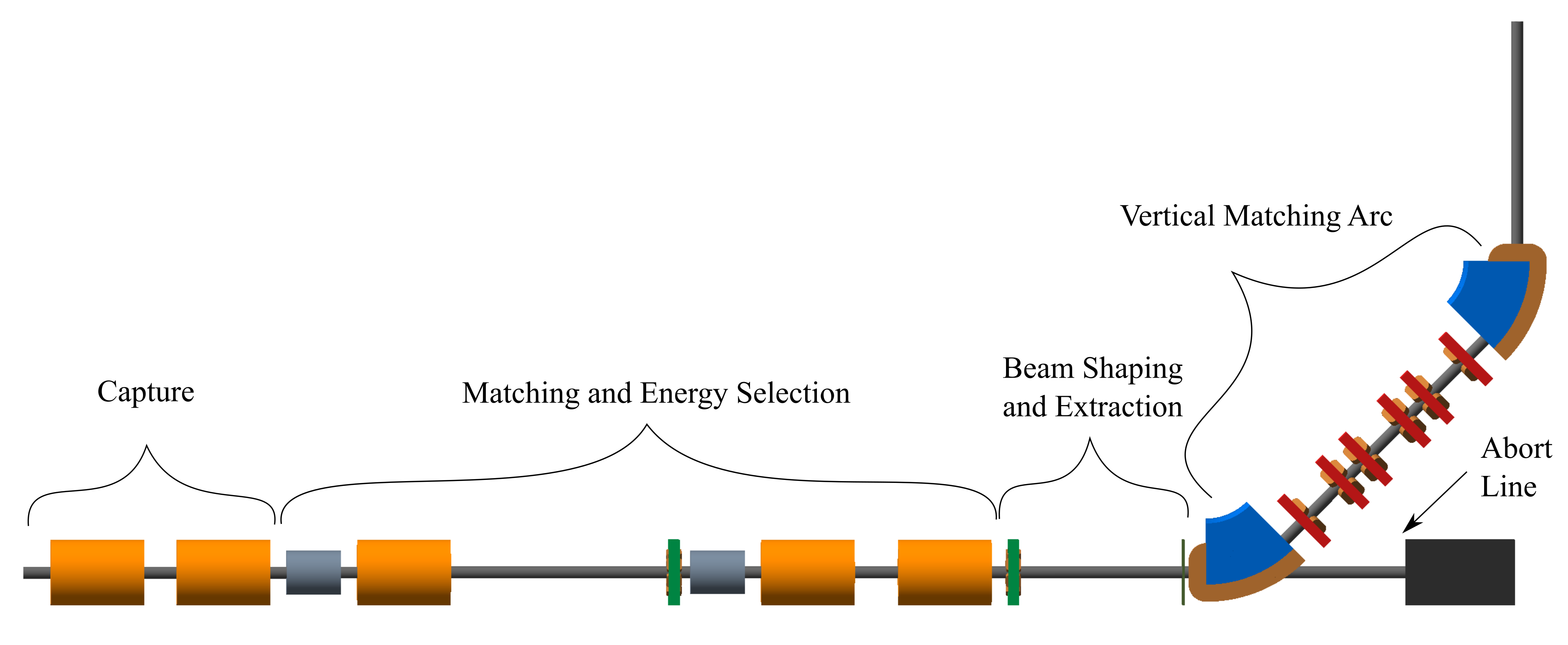}
	\end{center}
	\caption{ Beam transport for Stage 1 of LhARA visualised in BDSIM, showing five machine sections. The capture section is composed of two Gabor lenses (orange cylinders). The matching and energy selection section includes three Gabor lenses, two RF cavities (grey cylinders) and an octupole magnet (green disc). The beam shaping and extraction section includes a second octupole and a collimator (black vertical bar). The vertical matching arc directs the beam into the low-energy \emph{in vitro} end station and is composed of two 45$^\circ$ dipoles and six quadrupoles. The total length of this beam line is 17.255\,m. } 
	\label{fig:stage1overview}
\end{figure}

The beam line, shown schematically in figure \ref{fig:stage1overview},
is composed of five sections: beam capture; matching and energy
selection; beam shaping; vertical arc matching; and an abort line.
The capture section uses two Gabor lenses to minimise the transverse 
momentum of particles in the beam.
Beyond the capture section, an RF cavity permits control of the bunch
length and manipulation of the longitudinal phase-space.
A third Gabor lens then focuses the bunch to a small spot size after
which a second RF cavity is located to provide further longitudinal
phase-space manipulation.
Two further Gabor lenses bring the beam parallel once more in
preparation for the vertical 90$^\circ$ arc.
All Gabor lenses have an inner radius of 3.65\,cm and an effective
length of 0.857\,m. All lenses operate below the maximum cathode voltage
of 65\,kV.

A parallel beam emerges from the final Gabor lens, providing
significant flexibility for the inclusion of beam shaping and
extraction systems.
Beam uniformity will be achieved using octupole magnets to provide
third-order focusing to perturb the first-order focusing from
the Gabor lenses.
Such schemes have been demonstrated in a number of
facilities~\cite{beamuniformity1,beamuniformity2,beamuniformity3}.
A suitable position for the first octupole was identified to be after
the final Gabor lens where the beam is large; its effect on the beam
is expected to be significant.
Octupoles were only modelled in BDSIM as GPT does not have
a standard component with an octupolar field.
The typical rectangular transverse distribution resulting from
octupolar focusing requires collimation to match the circular 
aperture through which the beam enters the end station.
A collimator is therefore positioned at the start of the vertical
arc.
Further simulations are required to determine the optimum position of
the second octupole and to evaluate the performance of the octopoles.
The switching dipole which directs the beam to the injection line of
the FFA in Stage~2 will be located between the second octupole and the
collimator, requiring the octupole to be ramped down for Stage~2
operation. 

The vertical arc uses transparent optics in an achromat matching
section to ensure that the first-order transfer map through the arc is
equivalent to the identity transformation and that any dispersive
effects are cancelled.
A 2\,m drift tube is added after the arc to penetrate the concrete
shielding of the end station floor and to bring the beam to bench
height.
The abort line consists of a drift followed by a beam dump and
requires the first vertical dipole to ramp down, preventing
charged-particle transportation to the end station. 

\begin{figure}
	\begin{center}
		\includegraphics[width=0.9\textwidth]{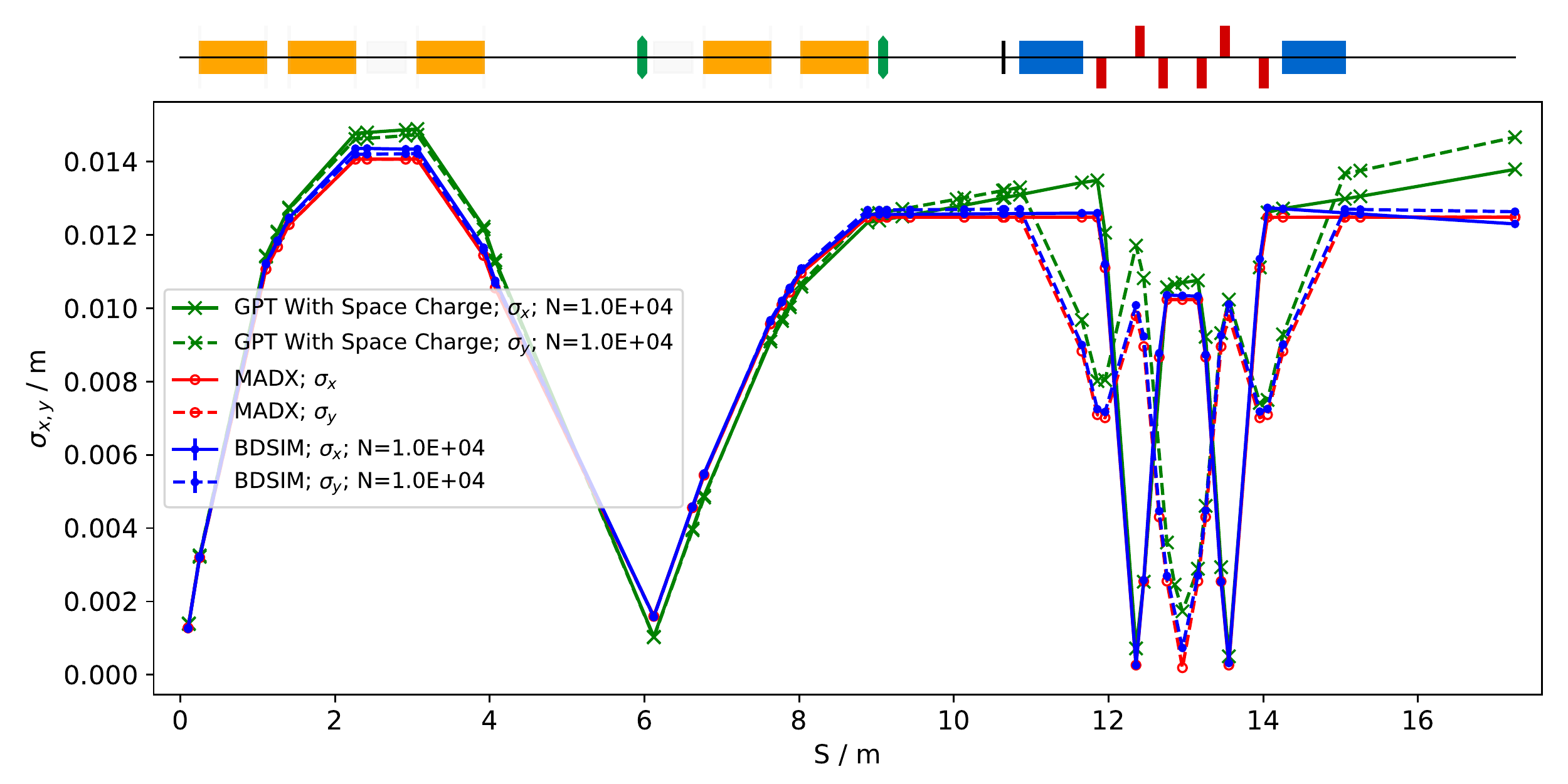}
	\end{center}
	\caption{Horizontal (solid lines) and vertical (dashed lines) beam sizes through the \emph{in vitro} beam transport, simulated with space-charge in GPT (green), and without space-charge in MADX (red) and BDSIM (blue).} 
	\label{fig:stage1optics}
\end{figure}

The underlying physics of plasma-lens operation cannot be simulated
in BDSIM or GPT, however it can be approximated using solenoid magnets
of equivalent strength. 
RF cavity fields were not simulated.
10\,000 particles were simulated corresponding to the estimated maximum
bunch charge of $1\times 10^{9}$ protons.
Figure~\ref{fig:stage1optics} shows excellent agreement between
horizontal and vertical transverse beam sizes in BDSIM and MADX,
verifying the beam line's performance in the absence of space-charge
effects. 
Reasonable agreement between BDSIM and GPT is also seen when
space-charge forces are included in GPT.
Emittance growth is observed prior to the first solenoid, affecting
the optical parameters throughout the machine.
However, the resulting beam dimensions at the cell layer of 1.38\,cm
horizontally and 1.47\,cm vertically are not significantly different
from the ideal beam in BDSIM.
Further adjustments of the Gabor lenses and arc-quadrupole strengths
may compensate for this.
The transmission efficiency of the beam line is approximately 100\%.

The small bunch dimensions in both transverse planes at the focus
after the third Gabor lens, where the energy selection collimator will
be placed, remains a concern if the effect of space-charge has been
underestimated.
Similar bunch dimensions are achieved in the vertical arc, however,
quadrupolar focusing is confined to a single plane mitigating further
emittance growth.
Further tuning of the Gabor lens voltages in the capture section may
compensate space-charge effects, reducing the non-zero transverse
momentum seen entering the vertical arc. 

\begin{figure}
	\centering     
	\includegraphics[width=0.99\textwidth]{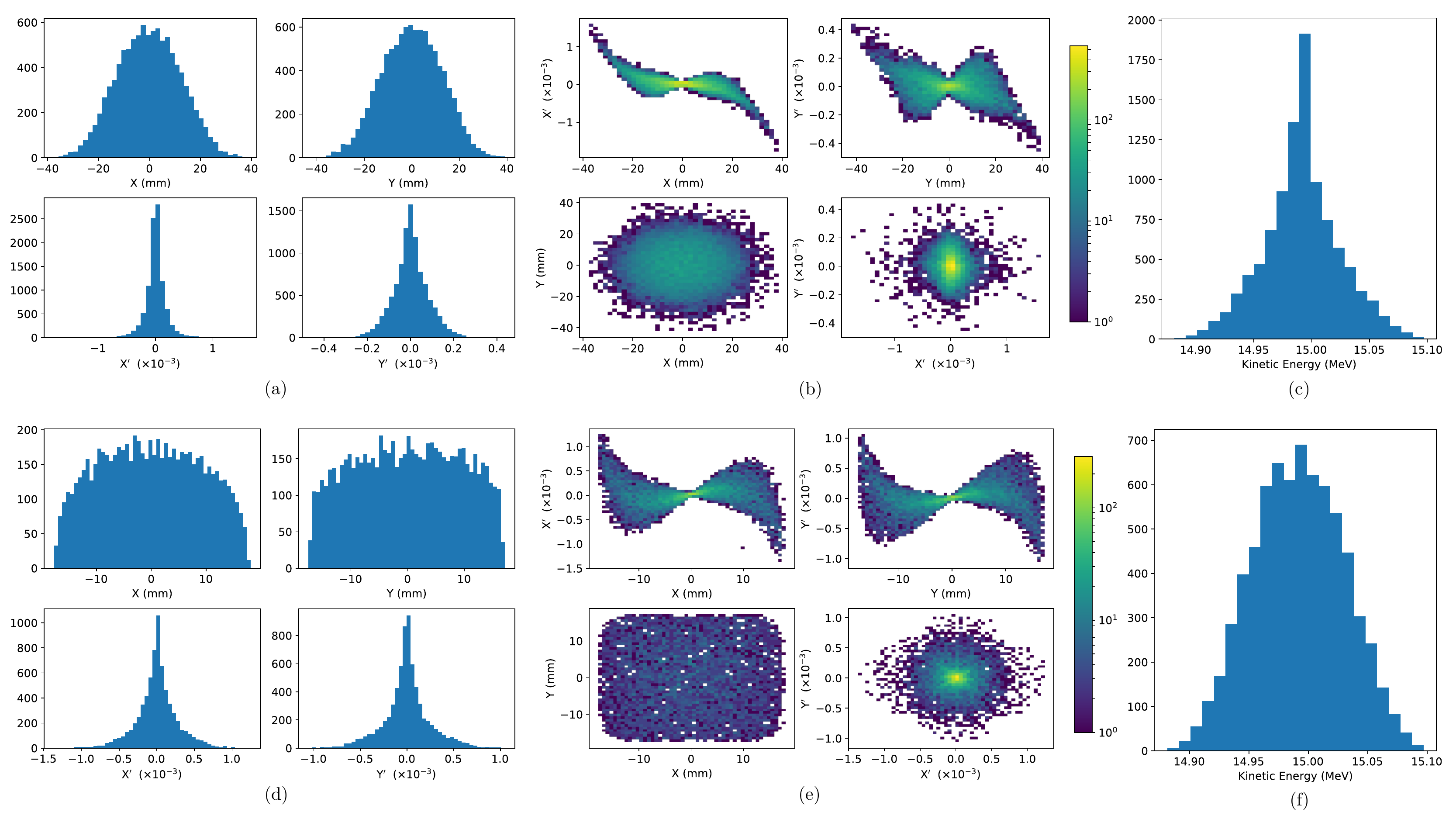}
	\caption{Beam phase space distributions at the end-station in the transverse plane, (X, Y ); X$^\prime$ and Y$^\prime$ give the slope relative to the Z axis. The transverse phase space is shown in figures a and b for simulations without octupolar focusing and collimation, with the kinetic energy distribution shown in c. The same phase space distributions simulated with the effect of octupoles and collimation are in figures d, e, and f.}
	\label{fig:phasespace}
\end{figure}

To investigate beam uniformity, BDSIM simulations with and without
octupoles and collimation for beam shaping were conducted.
Both octupoles were arbitrarily set to a strength of $\rm{K}3=6000$
with a magnetic length of 0.1\,m and pole-tip radius of 5\,cm, which,
for a 15\,MeV beam corresponds to pole-tip field of 0.42\,T.
A 2\,cm thick iron collimator with a 40\,mm diameter aperture was
positioned 1.5\,m downstream of the octupole.
Figure~\ref{fig:phasespace} shows the beam phase-space and particle
distributions at the end station for the transverse and longitudinal
axes with and without beam shaping.
Without octupoles the spatial profile is Gaussian as expected,
however, beam uniformity is improved with octupoles and collimation.
The total beam width is 3.58\,cm horizontally and 3.46\,cm vertically
which is sufficient to irradiate one well in a six-well cell-culture
plate.
Further optimisation is required to improve uniformity whilst
optimising beam-line transmission, which is approximately 70\% for the
results presented in figure~\ref{fig:phasespace}.
An aberration can be seen in both transverse planes with and without
beam shaping; this effect originates upstream of the octupoles in the
solenoids, and persists through to the end station. 
These aberrations are a concern, however, future simulation efforts
will replace the solenoids with a full electromagnetic simulation of
the Gabor lens.
This change is likely to change the aberrations.
The non-Gaussian energy distribution without beam shaping is a result
of space-charge forces at the ion source; the distribution persists to
the end station as no components which affect the longitudinal phase
space were simulated.
The Gaussian distribution seen with beam shaping is due to collimation.

The proposed design is capable of delivering beams of the desired size
to the \emph{in vitro} end station.
Space-charge effects impact the beam-transport performance but it is
believed that this can be mitigated with minor adjustments to the
Gabor lenses in the capture section.
Initial studies indicate that a uniform beam can be delivered with
further optimisation of the octupoles and collimator. \\

\subsubsection{Alternative Design}

\noindent To mitigate potential emittance growth from space-charge forces, an
alternative beam line design was developed in which the final two Gabor
lenses in the matching and energy selection section are replaced by
four quadrupoles, limiting any bunch focusing to one plane at a time.
The resulting machine is reduced in length to 15.439\,m.
Without space-charge effects, a beam sigma of 2.5\,mm at the end
station can be achieved.
With space-charge, emittance growth prior to the first
solenoid is once again observed leading to an increased beam size at
the entrance of the first quadrupole, resulting in a spatially
asymmetric and divergent beam at the end station.
It is believed that the  space-charge effects can be compensated by
applying the same Gabor-lens optimisation as in the baseline design and
adjusting the quadrupole settings to deliver beam parameters similar to
those without achieved in the absence of space charge.
The alternative design provides a solution that is more resilient to
space-charge effects than the baseline, however, only the lower bound
on the desired beam size has been achieved so far.
Further optimisation is required not only to optimise optical
performance but also to optimise octupole settings and to determine
whether a beam with the desired uniformity can be delivered to the end
station.

\graphicspath{ {03-LhARA-facility/03-04-Post-acceleration/Figures/} }

\subsection{Post-acceleration and beam delivery to the \emph{in vitro} and
  \emph{in vivo} end stations}
\label{SubSect:LhARA:invivoBeam}

\noindent A fixed-field alternating-gradient accelerator (FFA), based on
the spiral scaling principle \cite{ffa1, ffa2, ffa3, ffa4}, will be
used to accelerate the beam in LhARA Stage~2 to obtain energies
greater than the 15\,MeV protons and 4\,MeV/u carbon (C$^{6+}$) ions
delivered by the laser-driven source.
FFAs have many advantages for both medical and radiobiological
applications such as: the capability to deliver high and variable
dose; rapid cycling with repetition rates ranging from 10\,Hz to
100\,Hz or beyond; and the ability to deliver various beam energies
without the use of energy degraders.
An FFA is relatively compact due to the use of combined function
magnets, which lowers the overall cost compared to conventional
accelerators capable of delivering beams at a variety of energies such
as synchrotrons. 
Extraction can be both simple and efficient and it is possible for 
multiple extraction ports to be provided.
Furthermore, FFAs can accelerate multiple ion species, which is very
important for radiobiological experiments and typically very difficult
to achieve with cyclotrons. 

A typical FFA is able to increase the beam momentum by a factor of
three, though a greater factor may be achieved.
For LhARA, this translates to a maximum proton-beam energy of 127\,MeV
from an injected beam of 15\,MeV.
For carbon ions (C$^{6+}$) with the same rigidity, a maximum energy of
approximately 33.4\,MeV/u can be produced.  

The energy at injection into the FFA determines the beam energy at
extraction.
The injection energy will be changed by varying the focusing strengths
in the Stage~1 beam line from the capture section through to the
extraction line and the FFA ring. 
This will allow the appropriate energy slice from the broad energy
spectrum produced at the laser-driven source to be captured and
transported to the FFA.
The FFA will then accelerate the beam, acting as a three-fold momentum
multiplier.
This scheme simplifies the injection and extraction systems since
their geometry and location can be kept constant.

A second, `high-energy', \emph{in vitro} end station will be served by proton
beams with a kinetic energy in the range 15--127\,MeV and carbon-ion
beams with  energies up to 33.4\,MeV/u.
The extraction line from the FFA leads to a 90$^{\circ}$ vertical arc
to send the beam to the high-energy \emph{in vitro} end station.
If the first dipole of the arc is not energised, beam will be sent to
the \emph{in vivo} end station. 
The extraction line of the FFA includes a switching dipole that will
send the beam to the high-energy-beam dump if it is not energised.
The detailed design of the high-energy abort line, taking into account
the requirement that stray radiation does not enter the end stations,
will be performed as part of the LhARA R\&D programme.

\subsubsection{Injection line}

\begin{figure}
  \begin{center}
    \includegraphics[width=0.8\textwidth]{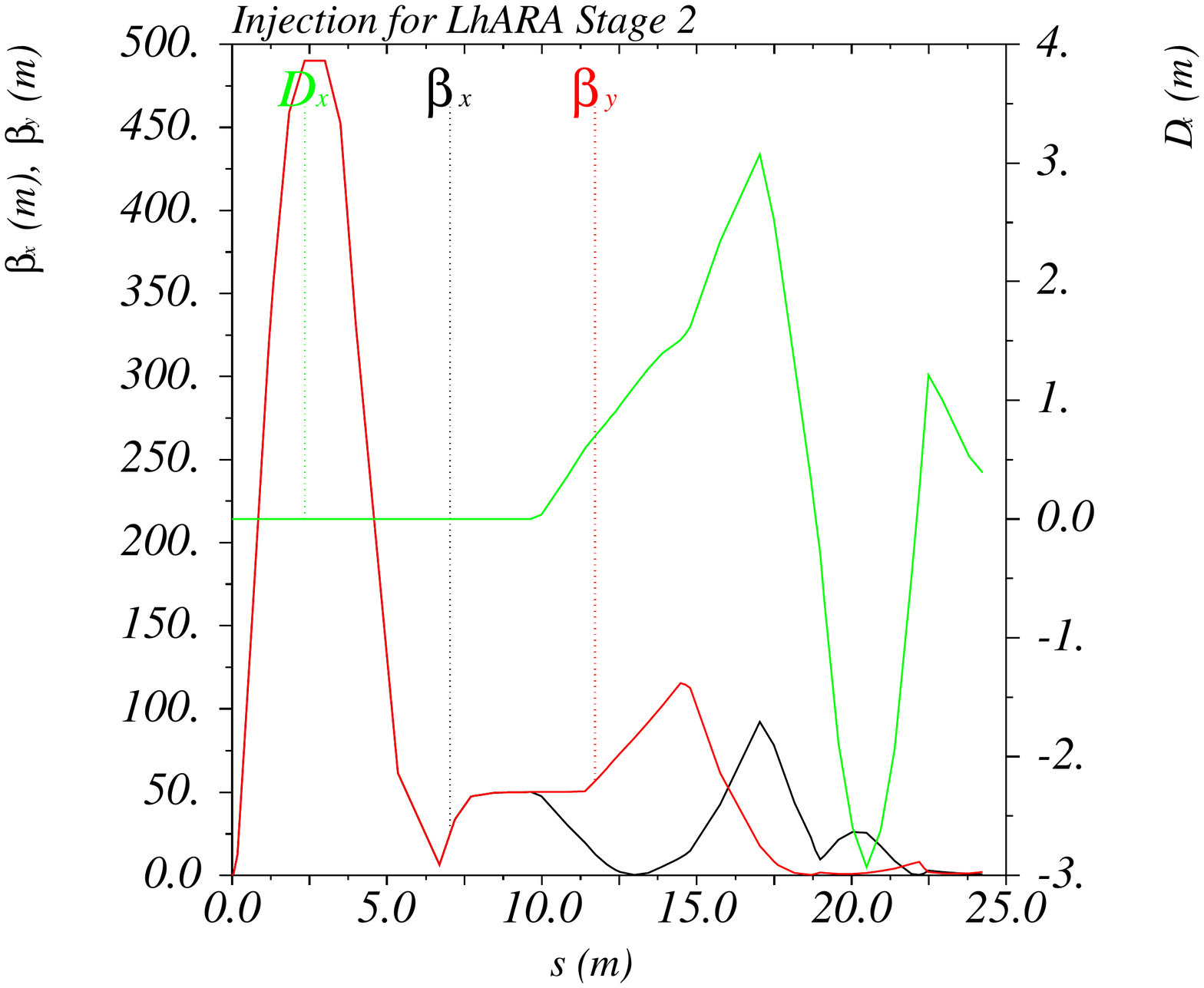}
  \end{center}
  \caption{
    Twiss $\beta_x$ and $\beta_y$ functions and dispersion in the beam line consisting of the modified Stage 1 lattice and the transfer line allowing injection of the beam into the FFA ring. $S$ goes from the laser target to the exit of the injection septum.
    }
  \label{fig:inj_Stage2}
\end{figure}

\begin{figure}
  \begin{center}
    \includegraphics[width=0.7\textwidth]{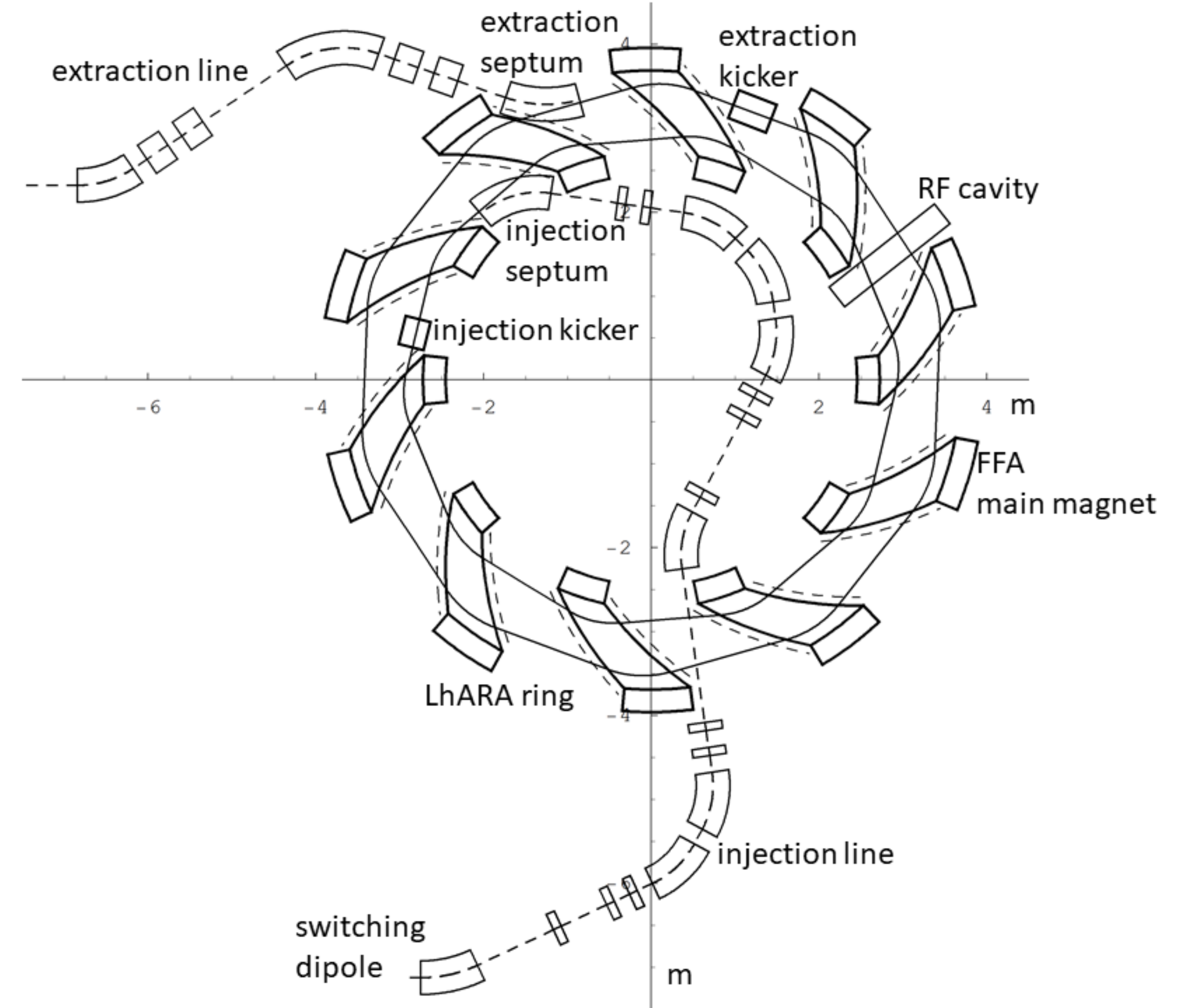}
  \end{center}
  \caption{
    The layout of the injection line from the switching dipole to the
    injection septum together with the FFA ring, some of its
    subsystems and the first part of the extraction line.
  }
  \label{fig:inj_ring}
\end{figure}

\noindent The settings of the Stage~1 beam line need to be adjusted to reduce
the Twiss $\beta$ function propagating through the injection line to
allow beam to be injected into the FFA ring.
The optical parameters in the Stage~1 beam line after adjustment are
shown in figure~\ref{fig:inj_Stage2}.
The beam is diverted by a switching dipole into the injection line which
transports the beam to the injection septum magnet.
The injection line matches the Twiss $\beta$ functions in both transverse planes and
the dispersion of the beam to the values dictated by the periodic conditions in
the FFA cell (figure~\ref{fig:inj_Stage2}).
The presence of dispersion in the injection line allows a collimator
to be installed for momentum selection before injection. 
The beam is injected from the inside of the ring, which requires the
injection line to cross one of the straight sections between the
FFA magnets, see figure~\ref{fig:inj_ring}.

\subsubsection{FFA ring}

\noindent The magnetic field, $B_y$, in the median plane of a scaling spiral FFA
is given by~\cite{ffa1, ffa2, ffa3}:
\begin{equation}
  B_y = B_0 \left[ \frac{R}{R_0} \right]^k
        F\left( \theta - \ln\left[ \frac{R}{R_0} \right] \tan\zeta  \right) \,;
  \label{Eq:FldBy}
\end{equation}
where $B_0$ is the magnetic field at radius $R_0$, $k$ is the field
index, $\zeta$ corresponds to the spiral angle and $F$ is the `flutter
function'. 
This field law defines a zero-chromaticity condition, which means the
working point of the machine is independent of energy up to field
errors and alignment imperfections. This avoids crossing any resonances, which
would reduce the beam quality and may lead to beam loss. 

Table~\ref{tab:parameters2} gives the main design parameters of the
FFA ring.
The ring consists of ten symmetric cells each containing a single
combined-function spiral magnet.
The choice of the number of cells is a compromise between the size of
the orbit excursion, which dictates the radial extent of the magnet,
and the length of the straight sections required to accommodate the
injection and extraction systems.

\begin{figure}
  \begin{center}
    \includegraphics[width=0.9\textwidth]{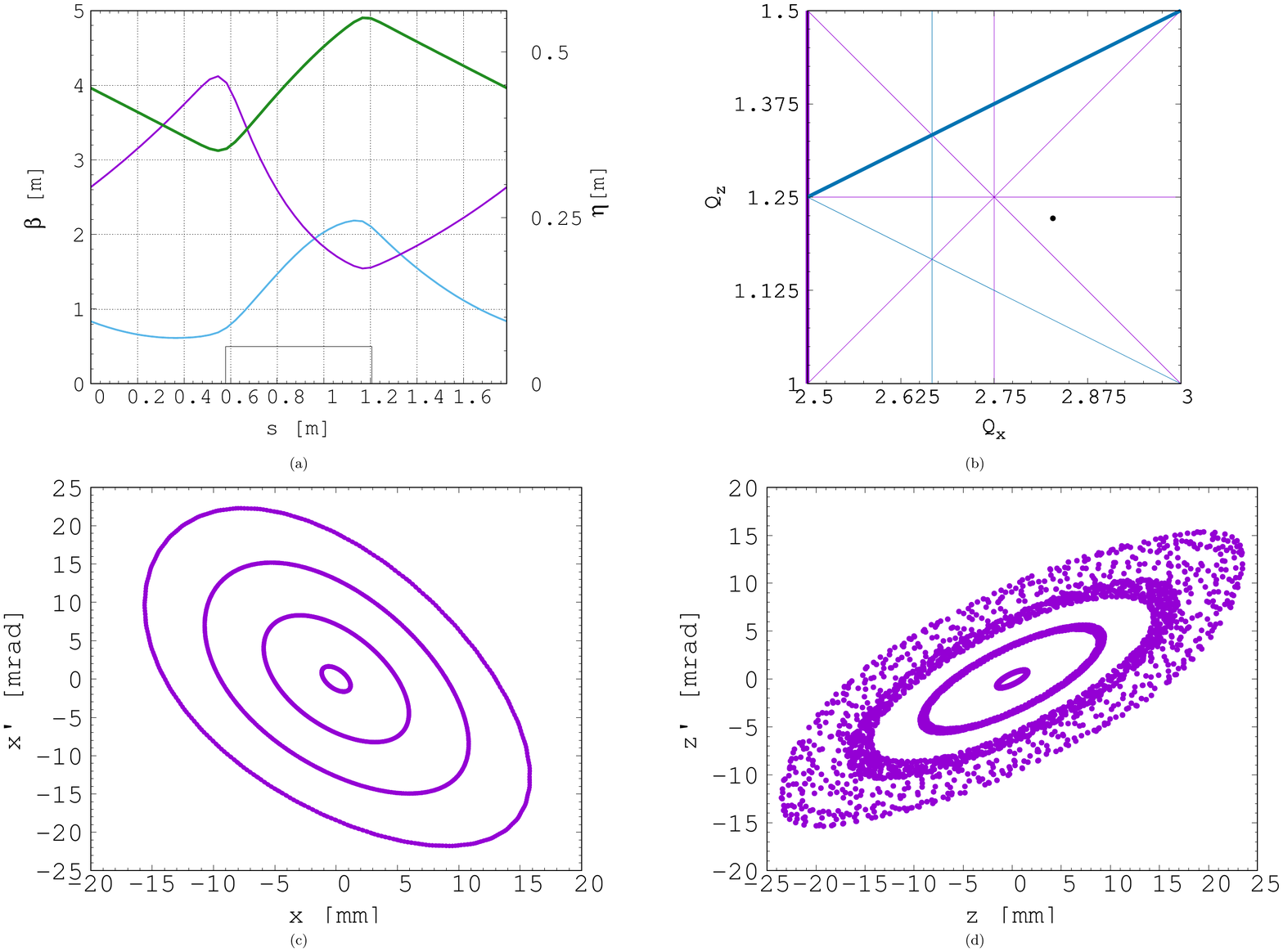}
  \end{center}
  \caption{
    Beam optics and tracking in the FFA. 
    Twiss $\beta_h$ (blue), $\beta_v$ (purple) functions and
    dispersion (green) in one lattice cell of the FFA ring (a).
    The working point of the FFA ring at (2.83,~1.22) on the tune
    diagram (b).
    The results of the horizontal (c) and vertical (d) dynamical acceptance study
    in the FFA ring, where a 1\,mm offset is assumed in the vertical and horizontal planes respectively.
  }
  \label{fig:ffaFigs}
\end{figure}

The betatron functions and dispersion in one lattice cell at injection
are shown in figure~\ref{fig:ffaFigs}a.
The tune diagram, showing the position of the working point of the
machine in relation to the main resonance lines, is shown in
figure~\ref{fig:ffaFigs}b.
Tracking studies were performed using a step-wise tracking code in
which the magnetic field is integrated using a Runge-Kutta algorithm
\cite{nustorm-triplet}.
The magnetic field in the median plane was obtained using the ideal
scaling law (equation~\ref{Eq:FldBy}) and using using Enge functions
to give the fringe fields.
The field out of the median plane was obtained using Maxwell's equations and 
a $6^{\rm th}$-order Taylor expansion of the field.
The dynamic acceptance for 100 turns, shown for the horizontal and
vertical planes in figures~\ref{fig:ffaFigs}c and \ref{fig:ffaFigs}d
respectively, are significantly larger than the beam emittance.
This statement holds even when the most pessimistic scenario, in which 
the emittance is assumed to be ten times larger than nominal, is used. 
These results confirm that a good machine working point has been
chosen.

A full aperture, fast injection of the beam will be performed using a
magnetic septum, installed on the inside of the ring, followed by a
kicker magnet situated in a consecutive lattice cell, as shown in
figure~\ref{fig:inj_ring}.  
The specifications of the injection system are dictated by the
parameters of the beam at injection, which are summarised for the
nominal proton beam in table~\ref{tab:ffa_beam}.
The beam at injection has a relatively small emittance and short bunch
length, which limits the intensity accepted by the ring due to the
space-charge effect.
An intensity of approximately $10^9$ protons will be accepted by the 
ring assuming the nominal beam parameters.
Space-charge effects will be severe immediately after injection, but 
will quickly be reduced due to the debunching of the beam.
Fast extraction of the beam over the full aperture will be performed
using a kicker magnet followed by a magnetic septum installed in a
consecutive lattice cell close to the extraction orbit.
\begin{table}
  \caption{
    Summary of the main parameters for the proton beam at the
    injection to the FFA ring.
    These parameters correspond to the nominal (maximum) acceleration
    mode of operation.
  }
  \label{tab:ffa_beam}
  \begin{center}
    \begin{tabular}{| l | c | c |}
      \hline
      Parameter & Unit & Value \\ \hline
      Beam energy & MeV & 15\\ 
      Total relative energy spread & \% & $ \pm 2$ \\
      Nominal physical RMS emittance (both planes) & $\pi$\,m\,rad & $4.1 \times 10^{-7}$ \\
      Incoherent space charge tune shift & & -0.8\\
      Bunching factor & & 0.023\\
      Total bunch length & ns & 8.1 \\
      Bunch intensity & & $10^9$ \\
      \hline 
    \end{tabular}
  \end{center}
\end{table}

Acceleration of the beam to 127\,MeV will be done using an RF system
operating at harmonic number $h=1$ with an RF frequency range from
2.89\,MHz to 6.48\,MHz. 
The RF voltage required for 10\,Hz operation is 0.5\,kV.
However, at such a low voltage the energy acceptance at injection 
will be limited to $\pm 0.7$\% so a voltage of 4\,kV is required 
to increase the energy acceptance to $\pm 2$\%.
This voltage can be achieved with one cavity \cite{ffa-RF},
two cavities are assumed to provide greater operational stability.
Normal conducting spiral-scaling FFA magnets, similar to the ones
needed for LhARA, have been constructed successfully
\cite{ffa4,raccam-magnet} using either distributed,
individually-powered coils on a flat pole piece or using a
conventional gap-shaping technique. 
For the LhARA FFA, we propose a variation of the coil-dominated design
recently proposed at the Rutherford Appleton Laboratory in R\&D
studies for the upgrade of the ISIS neutron and muon source.
In this case, the nominal scaling field is achieved using a
distribution of single-powered windings on a flat pole piece.
The parameter $k$ can then be tuned using up to three additional
independently-powered windings.
The extent of the fringe field across the radius of the magnet must be 
carefully controlled using a `field clamp' to achieve
zero-chromaticity.
An active clamp, in which additional windings are placed around
one end of the magnet, may be used to control the flutter function and
thereby vary independently the vertical tune of the FFA ring. 
The FFA is required to deliver beams over a range of energy; each 
energy requiring a particular setting for the ring magnets.
Therefore, a laminated magnet design may be required to reduce the time 
required to change the field.
The magnet gap of 4.7\,cm given in table~\ref{tab:parameters2} is
estimated assuming a flat-pole design for the magnet.
The details of the design will be addressed in as part of the LhARA
R\&D programme.

\subsubsection{Extraction Line}

\noindent Substantial margins in the beam parameters were assumed in the design
of the extraction line from the FFA due to uncertainties in the beam
distributions originating from: the Stage~1 beam transport; the FFA
injection line; and potential distortions introduced by the presence 
of space-charge effects during acceleration in the ring.
Therefore, the beam emittance was allowed, pessimistically, to be as
large as a factor of ten greater than in the nominal value, which
was derived assuming that the normalised emittance is conserved
from the source, through the Stage~1 beam line, and in the FFA ring.
In the nominal case, the physical emittance of the beam is affected by
adiabatic damping only.
Substantial flexibility in the optics of the extraction line is
required, as the extraction line must accommodate a wide spectrum of beam
conditions to serve the \emph{in vitro} and \emph{in vivo} end-stations. 

Detailed studies were carried out for proton beams with kinetic
energies of 40\,MeV and 127\,MeV.
Table \ref{tab:targetbeta} gives the Twiss $\beta$ values for
different beam sizes for the 40\,MeV and 127\,MeV proton-beam
scenarios assuming a Gaussian beam distribution.
The optics and geometric acceptance of the system is approximately the
same for the 40\,MeV and 127\,MeV beams.
This justified the working hypothesis that beam emittance is
approximately the same for both beam energies.
This assumption will be revised as soon as space-charge simulations
for the entire system are available. 
\begin{table}
  \caption{
    Beam emittance values and target $\beta$ values for different beam
    sizes for 40\,MeV and 127\,MeV beams.
    The beam size is taken to be four times the sigma of the
    transverse beam distribution.
  }
  \label{tab:targetbeta}
  \begin{center}
    \begin{tabular}{| c | c | c | c | }
      \hline
      & 40\,MeV protons & 127\,MeV protons & 127\,MeV protons\\
      & (Nominal) & (Nominal) & (Pessimistic) \\
      \hline
      RMS Emittance ($\epsilon_x$, $\epsilon_y$) [$\pi$ mm mrad] & 0.137 & 0.137 & 1.37 \\ 
      $\beta$ [m] for a 1\,mm spot size & 0.46 & 0.46 & 0.039 \\
      $\beta$ [m] for a 10\,mm spot size & 46 & 46 & 4.5 \\
      $\beta$ [m] for a 30\,mm spot size & 410 & 410 & 40 \\ 
      \hline
    \end{tabular}
  \end{center}
\end{table}

The first two dipoles and four quadrupoles of the extraction line
bend the beam coming from the extraction septum of the FFA such that
it is parallel to the low-energy beam line while ensuring that
dispersion is closed. 
Closing the dispersion is critical as off-momentum particles will
follow trajectories different to those followed by particles with the 
design momentum and therefore impact the size and shape of the beam
downstream.
The second part of the extraction line consists of four quadrupoles
which transport the beam either to the first dipole of the vertical
arc that serves the high-energy \emph{in vitro} end station or to the
\emph{in vivo} end-station if this dipole is not energised.
These quadrupoles provide the flexibility required to produce the
different beam sizes for the \emph{in vitro} end station as specified in
table~\ref{tab:targetbeta}.

\subsubsection{High-energy \emph{in vitro} beam line}

\noindent The high-energy \emph{in vitro} beam line transports the 
beam from the exit of the extraction line and delivers it to the 
high-energy \emph{in vitro} end station.
The 90$^{\circ}$ vertical bend is a scaled version of the low-energy
vertical arc, following the same design principles, and also consists
of two bending dipole magnets and six quadrupole magnets. 
To accommodate the higher beam energies, the lengths of the magnets
were scaled in order to ensure that peak magnetic fields were below
the saturation limits of normal conducting magnets.
The bending dipole magnet lengths were increased to 1.2\,m each and
the quadrupole lengths were tripled to 0.3\,m each.
The overall length of the arc then becomes 6\,m, compared to 4.6\,m
for the low energy \emph{in vitro} arc.
This difference in arc length means the high-energy \emph{in vitro} arc
finishes about 0.9\,m higher than the low-energy one.
This difference can easily be accommodated by adjusting the final
drift lengths.

\begin{figure}
  \begin{minipage}{.49\textwidth}
    \includegraphics[width=\textwidth]{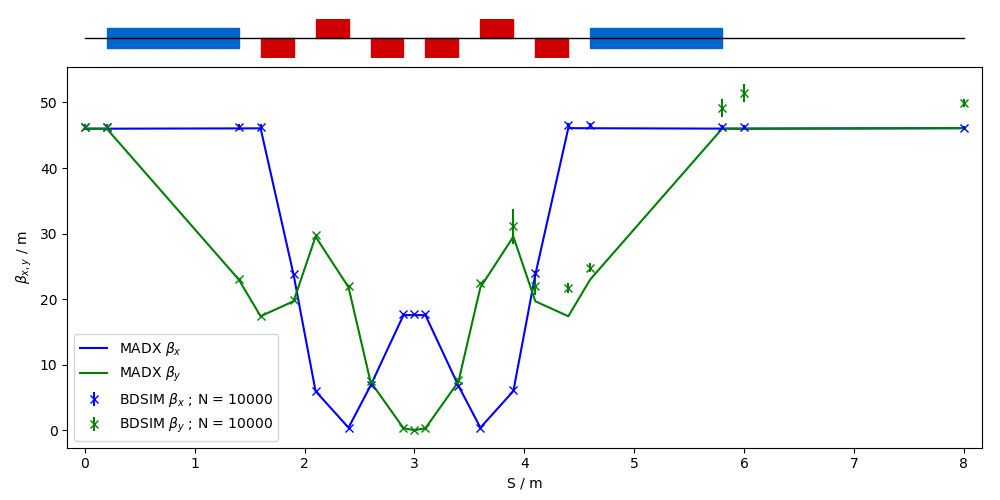}
  \end{minipage}
  \begin{minipage}{.49\textwidth}
    \includegraphics[width=\textwidth]{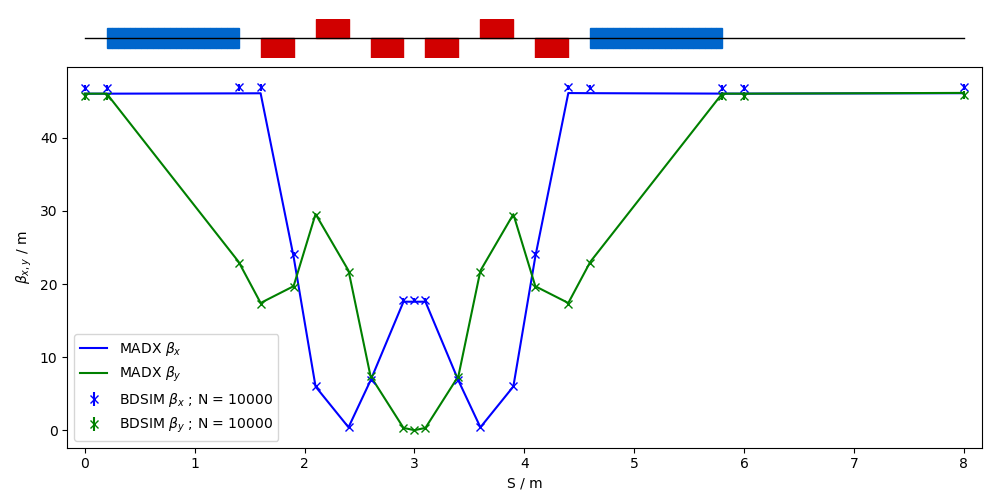}
  \end{minipage}
  \caption{
    Comparison of MAD-X and BDSIM simulation of 40\,MeV (left) and
    nominal 127\,MeV (right) proton beam passing through the high
    energy \emph{in vitro} arc simulated with $10^4$ particles (in BDSIM).
  }
  \label{fig:s2invitro_bdsim}
\end{figure}

The quadrupole strengths for the scaled high-energy \emph{in vitro} arc were
obtained using MAD-X and tracking simulations using BDSIM show good
agreement, see figure~\ref{fig:s2invitro_bdsim}.
The input beam distribution used in BDSIM was assumed to be Gaussian
with Twiss $\beta =46$, which gives a beam size of about 10\,mm.
GPT simulations were performed which show small discrepancies due to
space-charge effects.
It may be possible to compensate for this by adjusting the strengths
of the quadrupoles in the arc and the matching section in the
extraction line. \\

\subsubsection{\emph{In vivo} beam line}
\label{sec:s2invivo}

\noindent If the first dipole of the high-energy \emph{in vitro} arc is not energised
then beam is sent to the \emph{in vivo} end station.
From the end of the extraction line, 7.7\,m of drift is necessary to
clear the first bending dipole of the \emph{in vitro} arc, to provide space
for the five RF cavities needed for longitudinal phase-space
manipulation and to allow space for diagnostic devices.
Following this drift is a further 6.6\,m of beam line that includes
four quadrupoles, each of length 0.4\,m, which are used to perform the
final focusing adjustments of the beam delivered to the \emph{in vivo} end
station.
A final 1.5\,m drift at the end is reserved for scanning magnets to be
installed to perform spot scanning and to penetrate the shielding of
the \emph{in vivo} end station.
In total the \emph{in vivo} beam line is 15.6\,m in length.

\begin{figure}
  \begin{center}
  \begin{minipage}{.32\textwidth}
    \begin{center}
      \includegraphics[width=\textwidth]{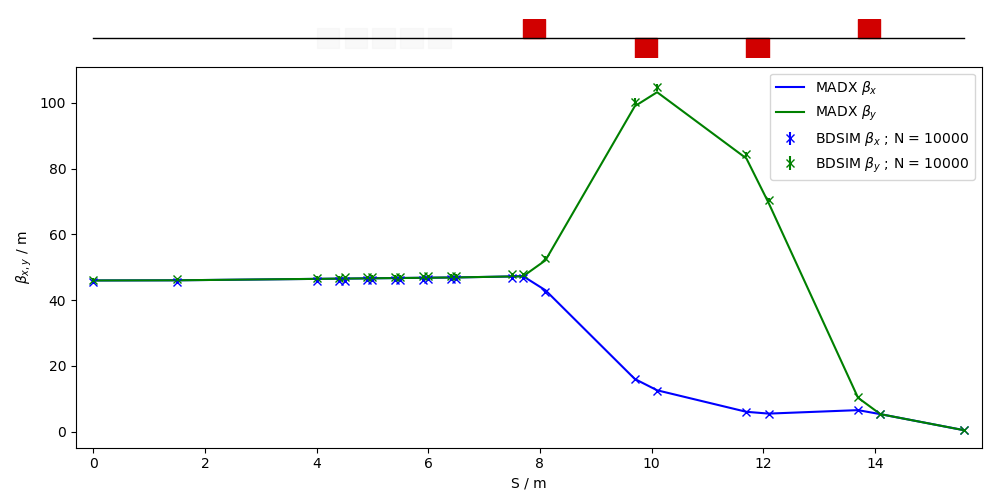}
    \end{center}
  \end{minipage}
  \begin{minipage}{.32\textwidth}
    \begin{center}
      \includegraphics[width=\textwidth]{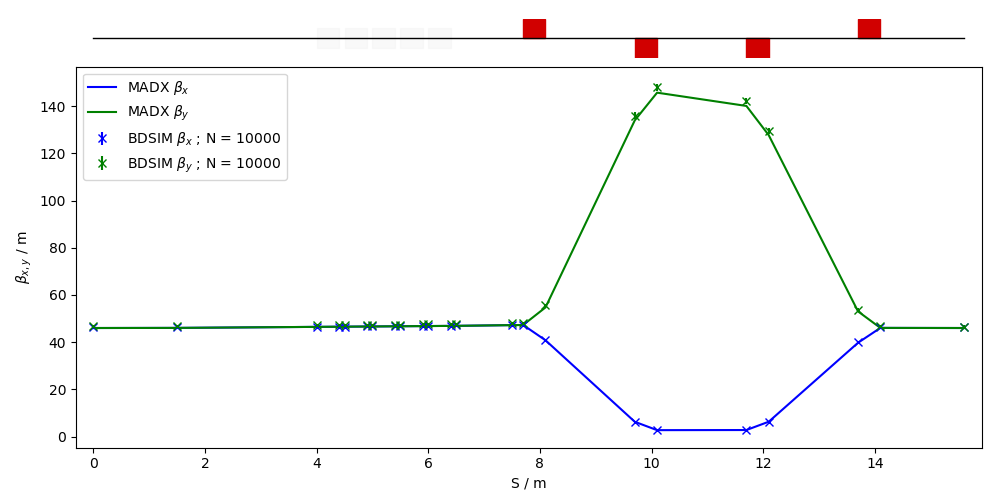}
    \end{center}
  \end{minipage}
  \begin{minipage}{.32\textwidth}
    \begin{center}
      \includegraphics[width=\textwidth]{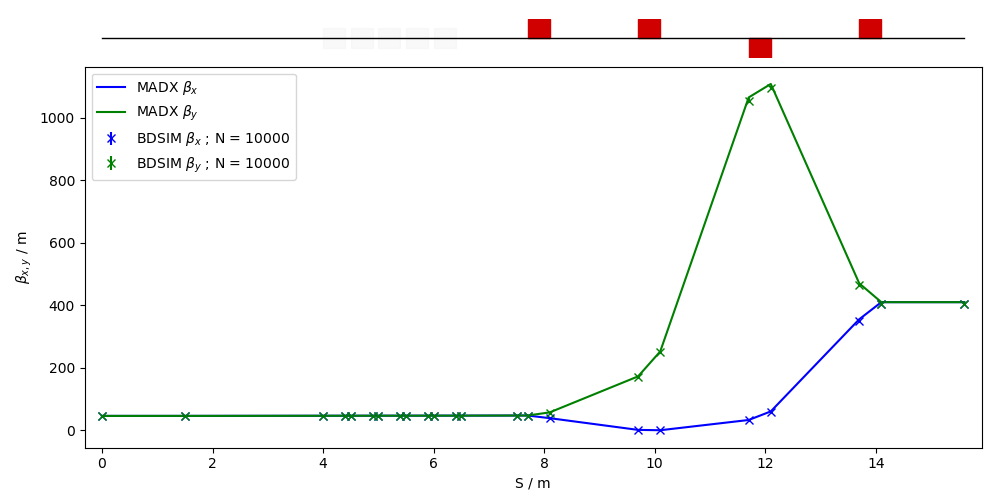}
    \end{center}
  \end{minipage}
  \end{center}
  \begin{center}
  \begin{minipage}{.32\textwidth}
    \begin{center}
      \includegraphics[width=\textwidth]{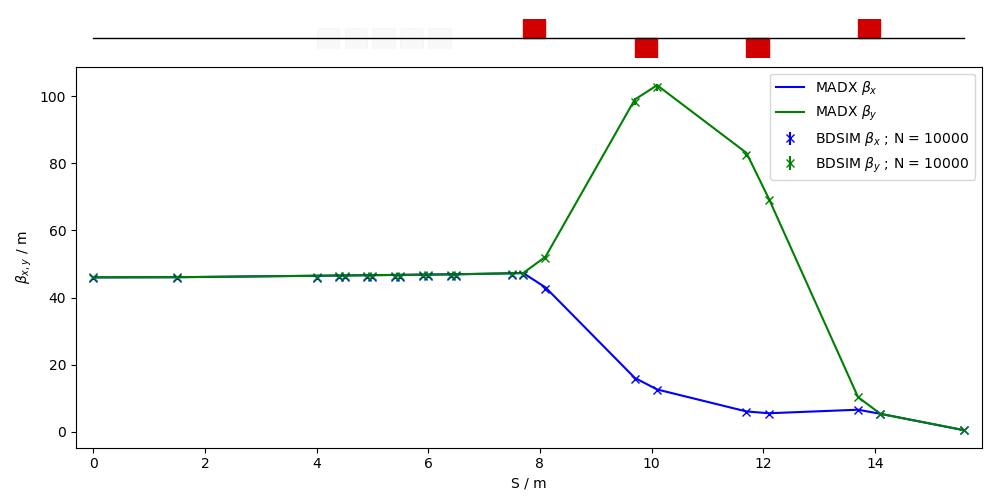}
    \end{center}
  \end{minipage}
  \begin{minipage}{.32\textwidth}
    \begin{center}
      \includegraphics[width=\textwidth]{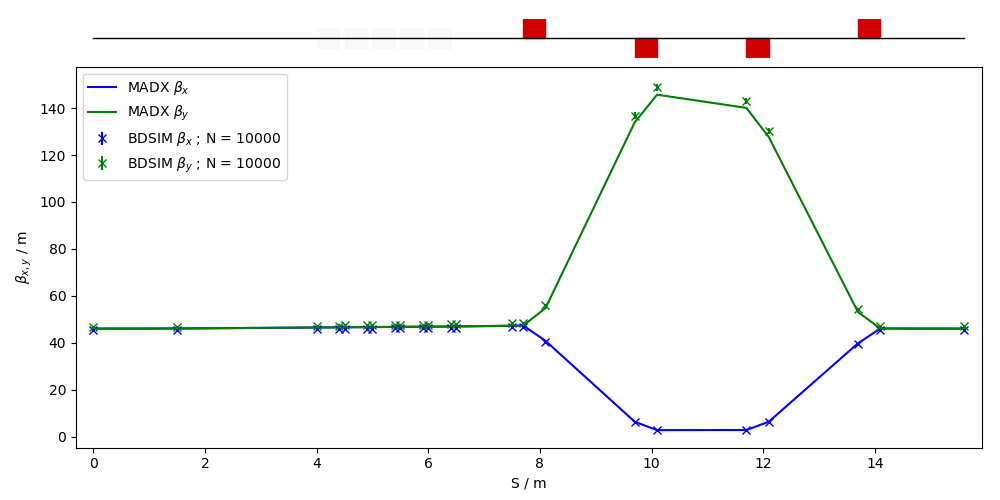}
    \end{center}
  \end{minipage}
  \begin{minipage}{.32\textwidth}
    \begin{center}
      \includegraphics[width=\textwidth]{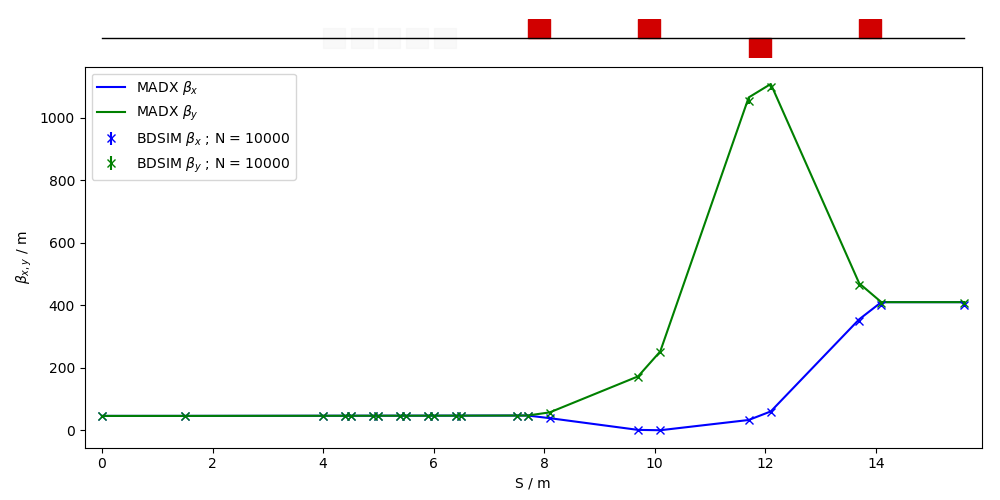}
    \end{center}
  \end{minipage}
  \end{center}
  \caption{
    MAD-X and BDSIM simulations of the \emph{in vivo} beam line for
    a 40\,MeV proton beam (top row) and a nominal 127\,MeV proton beam 
    (bottom row) with quadrupoles matched to
    $\beta_{x,y}=0.46$\,m (left), $\beta_{x,y}=46$\,m (middle) and $\beta_{x,y}=410$\,m (right) for $10^4$ particles.
  }
  \label{fig:invivo_bdsim_beta}
\end{figure}

The design is flexible in matching the various $\beta_{x,y}$ 
values given in table~\ref{tab:targetbeta}, but is not able to match
the smallest target value of $\beta_{x,y} = 0.039$\,m for the
pessimistic scenario, which is very challenging.
To verify that the optics design could provide the required beam
sizes, simulations were performed with BDSIM using an input Gaussian
beam generated with the Twiss $\beta$ values given in
tables~\ref{tab:targetbeta}.
Figure~\ref{fig:invivo_bdsim_beta} shows the results of simulations
for a 40\,MeV proton beam and a nominal emittance 127\,MeV proton beam
matched in order to obtain beam sizes of 1\,mm, 10\,mm and 30\,mm.
GPT was used to investigate the effects of space-charge.
These simulations show discrepancies compared to the BDSIM
simulations.
These discrepancies can be compensated for by adjusting the strengths
of the quadrupoles in the matching section in the extraction line. 

\graphicspath{ {03-LhARA-facility/03-06-Instrumentation/Figures/} }

\subsection{Instrumentation}
\label{SubSect:LhARAFac:Instr}

\noindent Commercial off-the-shelf (COTS) instrumentation will be used for
Stages~1 and 2 of LhARA wherever possible.
However, the characteristics of the beam (e.g. very high 
charge-per-bunch, low-to-moderate energy) will require some custom 
solutions to be developed.
The authors are developing two concepts, termed SciWire and
SmartPhantom, for the low- and high-energy \emph{in vitro} end 
stations respectively.
These detectors can also be used for beam diagnostics.
This new instrumentation may find application at other facilities.
Instrumentation for the detection of secondary particles arising from the
interaction of the beam with tissue is not discussed here but is an
important area that will be studied in the future. \\

\subsubsection{SciWire}

\noindent For the Stage~1 beam, the maximum proton energy is 15 MeV.
Shot-to-shot characterisation of the beam is essential and requires
the use of a very thin detector with a fast response. 
The SciWire \cite{SciWire} is being developed to provide energy and intensity profile 
measurements for low-energy ion beams. A single SciWire plane consists of two layers of 250\,$\mu$m
square-section scintillating fibres, with the fibre directions in the two
layers orthogonal to each other.
A series of back-to-back planes provides a homogeneous volume of scintillator. 
If there are enough planes to stop the beam, the depth of penetration will
allow the beam energy to be inferred.
This is obviously a destructive measurement so it is envisaged that this type of measurement 
would only be used when experiments are not running. A single plane, however, can be used for
2D beam-profile measurements at the same time that beam is delivered for experiments.
Detection of the light from SciWire fibres may be by CMOS camera, or
using photodiodes.
If the instrumentation is sufficiently fast, the SciWire can be used to
derive feedback signals for beam tuning. \\

\subsubsection{SmartPhantom}

\noindent To study in real time the dose profile of Stage~2 beams, the
SmartPhantom \cite{SmartPhantom} is being developed. 
This is a water-filled phantom, which is instrumented with planes of
scintillating fibres, by which to infer the dose distribution with
distance.
The detection elements of the SmartPhantom are 250\,$\mu$m diameter,
round scintillating fibres.
Each fibre station consist of two planes of fibres, in which the fibre
directions are orthogonal.
Five fibre stations are arranged in the phantom in front of the
cell-culture flask.
The fibres may be coupled to photodiodes, or a CMOS camera. 
  Simulations in GEANT4 are being used to develop analysis techniques by
which to predict the position of the Bragg peak shot-by-shot.
The beam profile and dose delivered can then be calculated in real
time.
The key emphasis is to be able to derive these parameters from 
shot-by-shot data, and not purely from simulations.  \\
  
\subsubsection{Beam line Instrumentation}
  
\noindent The instrumentation requirement begins with the Ti:Sapphire laser.
The laser focal spot will be characterised using a camera-based system
and high-speed wavefront measurements~\cite{Wavefront} from COTS
vendors.  
  
For the Stage~1 beam line, beam position monitors (BPMs) will be needed
for beam steering.
Because of the low beam energy, non-intercepting BPMs using capacitive
pickup buttons will be used.
Custom pickups will be needed to match the beam pipe geometry but COTS
electronics are available.
The beam current will be monitored near the end of each beam line,
using integrating current toroids (ICT), backed up with the option of
insertable multi-layer Faraday cups (MLFC) to give absolute beam
current and energy measurements.
Beam profiles could be measured by SEM grids on both Stage~1 and
Stage~2 beam lines.
For Stage~1, these monitors will be mounted on pneumatic actuators to
avoid scattering. 
Each end station could be equipped with insertable ``pepper-pot''
emittance monitors and a transverse deflection cavity with fluorescent
screen could be provided for bunch shape measurements.  
  
The BPMs on the FFA will require pickup designs suitable for the
unusual, wide and shallow, vacuum vessel. 
The FFA at the KURNS facility in Kyoto is of a similar
layout \cite{KURNS} and uses a kicker and capacitive pickup to
perform tune measurements in each transverse direction.
A minimum of one BPM every second cell will be used in the FFA so that
the beam orbit can be measured.
BPMs will also be required close to the injection and extraction
septa.
The BPM system may be able to use COTS electronics, but the pickups
will be based on the KURNS design of multiple electrodes arranged
across the vacuum vessel width. 
   
The DAQ needs to be able to store calibration data and apply
corrections in real time.
It is necessary to be able to find the beam centre from a profile,
even when the profile may be non-Gaussian and possibly asymmetric.
FPGAs can be used to perform fast fitting and pattern recognition of
beam profiles.
The instrumentation will be integrated with the accelerator control
system to be able to provide fast feedback and adjustment of the beam
parameters in real time.  

\graphicspath{ {03-LhARA-facility/03-08-End-stations/Figures/} }

\subsection{Biological end stations}
\label{SubSect:LhARA:BioEndStat}

\noindent In order to deliver a successful radiobiological research programme, high-end and fully equipped \emph{in vitro} and \emph{in vivo} end-stations will be housed within the LhARA facility. The two \emph{in vitro} end-stations (high and low energy) will contain vertically-delivered beam lines which will be used for the irradiation of 2D monolayer and 3D-cell systems (spheroids and patient-derived organoids) in culture. The beam line within the end-stations will be housed in sealed units that will be directly sourced with appropriate gases (carbon dioxide and nitrogen), allowing for the cells within culture plates to be incubated for a short time in stable conditions prior to and during irradiation. This will also enable the chamber to act, where necessary, as a hypoxia unit (0.1\%--5\% oxygen concentration). Furthermore, these sealed units will contain robotics to enable simple movement of the numerous cell culture plates housed within to be placed into and taken away from the beam. 

The \emph{in vitro} end-stations will be located within a research laboratory equipped with up-to-date and state-of-the-art facilities. This, coupled with two separate end-stations and multiple workspaces, will enable multiple groups of researchers to perform productive and high-quality biological research. The laboratory will include all the vital equipment for bench-top science, sample processing and analysis (e.g. refrigerated centrifuges and light/fluorescent microscopes), along with the equipment required for contaminant-free cell culture (e.g. humidified CO$_2$ cell culture incubators, Class II biological safety cabinets), and for the storage of biological samples and specimens (e.g. $-20^{\circ}$C and $-80^{\circ}$C freezers and fridges). The laboratory will also house an X-ray irradiator (allowing direct RBE comparisons between conventional photon irradiation, and the proton and carbon ions delivered by the accelerator), hypoxia chamber (for long-term hypoxia studies), a robotic workstation (handling and processing of large sample numbers, assisting in high-throughput screening experiments), and an ultra-pure-water delivery system. These facilities will enable a myriad of biological end-points to be investigated in both normal- and tumour-cell models not only from routine clonogenic survival and growth assays, but will expand significantly on more complex end-points (e.g. inflammation, angiogenesis, senescence and autophagy) as these experiments are difficult to perform at current clinical research beams due to limited time and facilities. 

The \emph{in vivo} end-station will be served with high-energy proton and carbon ions capable of penetrating deeper into tissues allowing the irradiation of whole animals. The ability to perform \emph{in vivo} pre-clinical studies is vital for the future effective translation of the research into human cancer patients where optimum treatment strategies and reduction of side-effects can be defined. The \emph{in vivo} end-station will allow the irradiation of a number of small-animal models (e.g. xenograft mouse and rat models) which can further promote an examination of particular ions on the appropriate biological end-points (e.g. tumour growth and normal tissue responses). The end-station will contain a small-animal handling area which will allow for the anaesthetisation of animals prior to irradiation. To enable the irradiation of small target volumes with a high level of precision and accuracy, an image guidance system (e.g. computed tomography) will be available. The animals will subsequently be placed in temperature-controlled holder tubes enabling the correct positioning of the relevant irradiation area in front of the beam line. The beam size is sufficient to give flexibility in the different irradiation conditions, in particular through passive scattering, pencil-beam scanning, and micro-beam irradiation, to be investigated at both conventional and FLASH dose rates. It is envisaged that the animals will be taken off-site post-irradiation to a nearby animal-holding facility for a follow-up period where biological measurements will be conducted.

\graphicspath{ {03-LhARA-facility/03-07-Infrastructure/Figures/} }

\subsection{Infrastructure and integration}
\label{SubSect:LhARA:InfInt}

\noindent The LhARA facility will encompass two floors of roughly 42\,m in length and 18\,m wide. The ground floor will contain the laser, accelerator, and \emph{in vivo} end station while the first floor will house the laboratory area and the two \emph{in vitro} end stations. The entire facility will require radiation protection in the form of concrete shielding, which will delineate the facility into three principal areas: a radiation-controlled access area, a laser controlled access area, and a laboratory limited-access area. 

It is envisaged that LhARA will be built at an STFC National Laboratory or equivalent research institute which has an established safety-management system and culture in place. At STFC, a comprehensive set of Safety Codes has been developed to cover the hazards associated with working in such an environment. STFC Safety Codes applicable to LhARA include: risk management, construction, biological safety, working with lasers, working with time-varying electro-magnetic fields, management of ionising radiation, and electrical safety. In practice at STFC, these codes are backed-up by the knowledge, skills and experience of staff, and by appointed responsible persons such as Radiation Protection Advisors, Laser Responsible Officers, and Authorising Engineers. In addition, STFC operates many facilities that encompass the same hazards as LhARA, which, for lasers, include the Gemini Target Areas 2 and 3 \cite{Gemini} as well as the new EPAC (Extreme Photonics Application Centre) \cite{EPAC} and for accelerators include FETS (Front End Test Stand) \cite{FETS}, and the ISIS Neutron and Muon Source \cite{ISIS}. Safety systems and equipment will be required for LhARA, which will include Class II biological safety cabinets for contaminant-free cell culture for \emph{in vitro} radiobiological experiments.

For a facility such as LhARA, radiation safety is a primary concern and all work will be completed under Regulation 8 of the Ionising Radiations Regulations 2017 (IRR17) \cite{IRR17}, which requires a radiation risk assessment before commencing a new work activity involving ionising radiation. 

The infrastructure and integration of the LhARA facility will require R\&D in four key areas: risk analysis (project risks), risk assessments (safety risks), radiation simulations, and controls development. The risk analysis will cover all aspects of the facility, such as funding and resource availability, not just technical risks. A safety-risk assessment will be performed to describe and control all potential safety risks in the facility. The safety-risk assessment will, to a reasonable degree, identify all pieces of equipment that require safety mitigations and identify control measures that must be put in place. Coupled closely with the safety-risk assessment, radiation simulations will be developed to characterise the radiation hazards in and around the LhARA facility. The last area to require R\&D will be the control systems. It is expected that the facility will use the Experimental Physics and Industrial Control System (EPICS), which can be further developed at this stage. \\

\graphicspath{ {04-Performance/Figures/} }

\section{Performance}
\begin{figure}
  \begin{center}
    \includegraphics[width=0.5\textwidth]{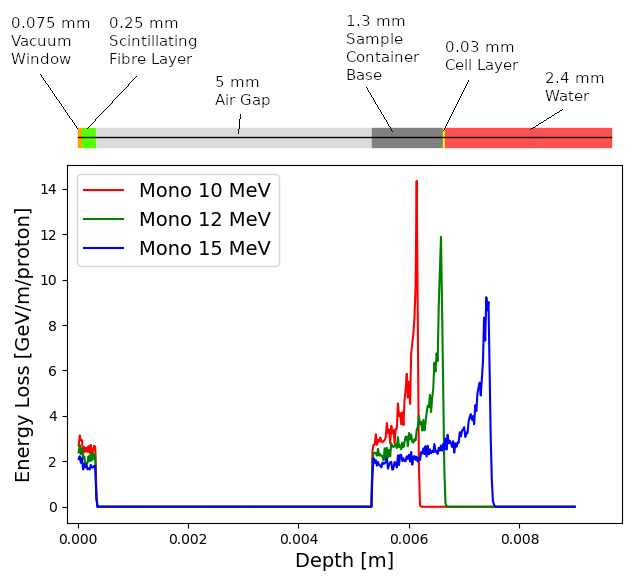}
  \end{center}
  \caption{
    Energy loss as a function of depth in the low-energy \emph{in vitro}
    end station for three mono-energetic proton energies: 10\,MeV;
    12\,MeV; and 15\,MeV.
    Each beam was simulated using $10^4$ particles at the start of the
    simulated end station.
  }  
  \label{fig:st1endstation}
\end{figure}

The dose distributions delivered to the end stations were evaluated
using BDSIM. 
Figure~\ref{fig:st1endstation} shows the energy lost by the beam as 
it enters the low-energy \emph{in vitro} end station.
The beam passes through the vacuum window, a layer of scintillating
fibre, and a 5\,mm air gap.
The beam then enters the cell-sample container, assumed to be
polystyrene, which supports a 30\,$\mu$m thick
layer of cells, modelled using the Geant4 material
``G4\_SKIN\_ICRP'' \cite{NIST}.
The transverse momentum of protons in the beam was assumed to be
Gaussian distributed, with a lateral spread small enough for the beam
to be fully contained within the required spot size of 3\,cm.
Figure~\ref{fig:st1endstation} shows that a proton beam with 10\,MeV
kinetic energy does not reach the cell.
The Bragg peak of a 12\,MeV proton beam is located close to the
cell layer, while a 15\,MeV beam, the maximum energy specified for
delivery to the low-energy \emph{in vitro} end station, has a Bragg peak 
located beyond the cell layer.
LhARA's ability to deliver various energies will allow the
investigation of radiobiological effects for irradiations using
different parts of the Bragg peak, effectively varying the LET across
the sample.

The maximum dose that can be delivered was evaluated for a variety of
beam energies. 
The dose was estimated from simulations by calculating the energy deposited 
in a volume of water corresponding in size to the sensitive volume of a 
PTW~23343 Markus ion chamber \cite{PTW_manual_detector} placed at the 
position of the Bragg peak in each case.
The cylindrical chamber has a radius of 2.65\,mm and a
depth of 2\,mm, giving a volume of about $4.4\times10^{-8}$\,m$^3$.
The total energy deposited within the chamber was recorded and converted
into dose in units of Gray.  

For the low-energy \emph{in vitro} end station the minimum spot size is
specified to have a diameter of 10\,mm, which is larger than the area
of the chamber.
A single shot of $10^9$ protons at 12\,MeV with the minimum design
spot size deposits $3.1 \times 10^{-4}$\,J in the chamber volume,
corresponding to a dose of $7.1$\,Gy.
For this simulation, the thickness of the sample container was reduced
so that the Bragg peak could be positioned within the chamber volume.
For the bunch length of $7.0$\,ns the maximum instantaneous dose rate is
$1.0\times10^{9}$\,Gy/s and the average dose rate is $71$\,Gy/s
assuming a repetition rate of 10\,Hz.
A single shot of $10^{9}$ protons at 15\,MeV deposits
$5.6 \times 10^{-4}$\,J in 
the chamber volume corresponding to a dose of $12.8$\,Gy.
This gives an instantaneous dose rate of $1.8\times10^{9}$\,Gy/s and
an average dose rate of $128$\,Gy/s assuming the same bunch length and 
repetition rate as for the 12\,MeV case.  

For the high-energy \emph{in vitro} end station a different setup was 
used for high energy proton beams. 
A similar design to the low-energy end station was used but with the
air gap increased from 5\,mm to 5\,cm and a water phantom was placed
at the end of the air gap instead of a cell culture plate.
The water phantom used in the simulation was based upon the PTC T41023
water phantom \cite{PTW_manual_phantom}.
In addition, the smaller minimum design beam size of 1\,mm was used. 
A single shot of $10^9$ protons at 127\,MeV deposits
$6.9 \times 10^{-4}$\,J in the chamber at the pristine Bragg peak
depth corresponding to a dose of $15.6$\,Gy, an instantaneous dose
rate of $3.8\times10^{8}$\,Gy/s and an average dose rate of
$156$\,Gy/s.
The end-station design assumed for a 33.4\,MeV/u carbon beam was the 
same as that used for the low-energy \emph{in vitro} end station due to 
the limited range in water of the carbon beam. 
The intensity of the beam is a factor of 12 less than for protons in
order to preserve the same strength of the space-charge effect at injection 
into the FFA with the same beam parameters, as the incoherent space charge 
tune shift is proportional to $q^2/A$ and inversely proportional to 
$\beta^2\gamma^3$, where $q$ corresponds to the particle charge, $A$ 
its mass number and $\beta$, $\gamma$ its relativistic parameters. 
A single pulse of $8.3\times10^{7}$ ions, deposits $3.2 \times 10^{-3}$\,J 
at the depth of the pristine Bragg peak, leading to an instantaneous dose 
rate of $9.7\times10^{8}$\,Gy/s and a maximum average dose rate of
$730$\,Gy/s.  

The expected maximum dose rates are summarised in 
table~\ref{tab:flash_lhara_doserates_3}.
The instantaneous dose rates depend on the bunch length which differs 
depending on the energies. 
For the low-energy \emph{in vitro} line a 7\,ns bunch length is assumed 
here for all energies. While for the higher energies, a 127\,MeV proton 
beam is delivered with a bunch length of 41.5\,ns, and a bunch length 
of 75.2\,ns for a 33.4\,MeV/u carbon beam. 
The same repetition rate of 10\,Hz was used for all energies.
The minimum beam size at the start of the end station for the 12\,MeV 
and 15\,MeV proton-beam simulations was 1\,cm. A 1\,mm beam size was 
used for the 127\,MeV proton beam and 33.4\,MeV/u carbon-ion beam 
simulations. 
\begin{table}
  \caption{
    Summary of expected maximum dose per pulse and dose rates that LhARA can
    deliver for minimum beam sizes.
    These estimates are based on Monte Carlo simulations using a bunch
    length of 7\,ns for 12\,MeV and 15\,MeV proton beams, 41.5\,ns for the 127\,MeV proton beam and 75.2\,ns for the 33.4\,MeV/u carbon beam. 
    The average dose rate is based on the 10\,Hz repetition rate of the laser
    source.
  }
  \label{tab:flash_lhara_doserates_3}
  \begin{center}
    \begin{tabular}{ | c | c | c | c | c | }
      \hline
       & 12\,MeV Protons & 15\,MeV Protons & 127\,MeV Protons & 33.4\,MeV/u Carbon \\
      \hline
      Dose per pulse & $7.1$\,Gy & $12.8$\,Gy & $15.6$\,Gy & $73.0$\,Gy \\
      \hline
      Instantaneous dose rate & $1.0\times10^{9}$\,Gy/s &
      $1.8\times10^{9}$\,Gy/s & $3.8\times10^{8}$\,Gy/s &
      $9.7\times10^{8}$\,Gy/s \\  
     \hline
      Average dose rate & $71$\,Gy/s & $128$\,Gy/s & $156$\,Gy/s & $730$\,Gy/s \\
      \hline
    \end{tabular}
  \end{center}
\end{table}

\graphicspath{ {05-Conclusions/Figures/} }

\section{Conclusions}

The initial conceptual design of LhARA, the Laser-hybrid Accelerator
for Radiobiological Applications, has been described and its 
performance evaluated in simulations that take into account the key
features of the facility.
LhARA combines a laser-driven source to create a large flux of protons
or light ions which are captured and formed into a beam by
strong-focusing plasma lenses thus evading the current space-charge
limit on the instantaneous dose rate that can be delivered.
Acceleration, performed using an fixed-field alternating-gradient
accelerator, preserves the unique flexibility in the time, spectral,
and spatial structure of the beam afforded by the laser-driven source.
The ability to trigger the laser pulse that initiates the production of
protons or ions at LhARA will allow the time structure of the beam to
be varied to interrupt the chemical and biological pathways that
determine the biological response to ionising radiation.
In addition, the almost parallel beam that LhARA will deliver can be
varied to illuminate a circular area with a maximum diameter of 
between 1\,cm and 3\,cm with an almost uniform dose or focused to a
spot with diameter of $\sim 1$\,mm.
These features make LhARA the ideally flexible tool for the systematic
study of the radiobiology of proton and ion beams.

The laser-hybrid approach, therefore, will allow radiobiological
studies and eventually radiotherapy to be carried out in completely
new regimes, delivering a variety of ion species in a broad range of
time structures and spatial configurations at instantaneous dose
rates up to and potentially significantly beyond the current
ultra-high dose-rate ``FLASH'' regime.
By demonstrating a triggerable system that incorporates
dose-deposition imaging in the fast feedback-and-control system, LhARA
has the potential to lay the foundations for ``best in class''
treatments to be made available to the many by reducing the footprint
of future particle-beam therapy systems. 

LhARA has the potential to drive a change in clinical practice in the
medium term by increasing the wealth of radiobiological knowledge.
This enhanced understanding in turn may be used to devise new 
approaches to decrease radio-toxicity on normal tissue while
maintaining, or even enhancing, the tumour-kill probability.
The radiobiology programme in combination with the demonstration in 
operation of the laser-hybrid technique means that the execution of the
LhARA programme has the potential to drive a step-change in the 
clinical practice of proton- and ion-beam therapy.

\section*{Acknowledgements}

The work described here was made possible by a grant from the Science
and Technology Facilities Council (ST/T002638/1, ST/P002021/1).
Additional support was provided by the STFC Rutherford Appleton and
Daresbury Laboratories and members of the LhARA consortium.
We gratefully acknowledge all sources of support.
A pre-publication review of the pre-CDR for LhARA was carried out by
P.~Bolton (LMU, Munich), M.~Lamont (CERN), Y.~Prezado (Institut Curie),
and F.~Romano (INFN-LNS and the National Physical Laboratory).
We are grateful to the review panel for their support and detailed
feedback on the draft pre-CDR.

\bibliographystyle{99-Styles/utphys}
\bibliography{bibliography}

\end{document}